\documentclass[11pt,reqno,oneside]{amsart} 

\usepackage{multicol}
\usepackage{graphicx}
\usepackage{amsmath}
\usepackage{amssymb}
\usepackage{algpseudocode}
\usepackage{xcolor}
\usepackage{fancyhdr}
\usepackage{mathrsfs}
\usepackage{mathabx}
\usepackage[margin=1in,bottom=.65in]{geometry} 
\usepackage{float}
\usepackage{algorithm}
\usepackage{algpseudocode}
\usepackage{booktabs}
\usepackage{tikz}
\usepackage{comment}
\usepackage{subcaption}
\usepackage{float}
\usepackage[normalem]{ulem}
\usepackage[colorlinks=true,
            linkcolor=blue,
            citecolor=blue,
            urlcolor=blue]{hyperref}

\usetikzlibrary{arrows.meta, positioning, calc}
\makeatletter
\newcounter{figtab}

\let\c@figure\c@figtab
\let\c@table\c@figtab

\makeatother

\newcommand{\sech}{\operatorname{sech}}
\newtheorem{theorem}{Theorem}

\theoremstyle{remark}
\newtheorem{Remark}[theorem]{Remark}

\title[Solitons: classical methods vs NN schemes]{
Soliton profiles: Classical Numerical\\ Schemes vs. Neural Network\,-\,Based Solvers}

\author[C. Haight]{Chandler Haight}
\address{Department of Mathematics \&  Statistics, Florida International University, Miami, FL 33199, USA}
\email{chaig004@fiu.edu}

\author[S. Roudenko]{Svetlana Roudenko}
\address{Department of Mathematics \&  Statistics, Florida International University, Miami, FL 33199, USA}
\email{sroudenko@fiu.edu}

\author[Z. Wang]{Zhongming Wang}
\address{Department of Mathematics \&  Statistics, Florida International University, Miami, FL 33199, USA}
\email{zwang6@fiu.edu}

\date{}


\begin{document}

\begin{abstract} 


We present a comparative study of classical numerical solvers, such as Petviashvili's method or finite difference with Newton iterations, and  neural network-based methods for computing ground states or profiles of solitary-wave solutions to the one-dimensional dispersive PDEs that include the nonlinear Schr\"odinger, the nonlinear Klein-Gordon and the generalized KdV equations. We confirm that classical approaches retain high-order accuracy and strong computational efficiency for single-instance problems in the one-dimensional setting. Physics-informed neural networks (PINNs) are also able to reproduce qualitative solutions but are generally 
less accurate and less efficient in low dimensions than classical solvers due to expensive training and slow convergence. We also investigate the operator-learning methods, which, 
although computationally intensive during training, can be reused across many parameter instances, providing rapid inference after pretraining, making them attractive for applications involving repeated simulations or real-time predictions. For single-instance computations, however, the accuracy of operator-learning methods remains lower than that of classical methods or PINNs, in general. 
\end{abstract}

\maketitle

\section{Introduction}
Nonlinear dispersive equations such as the nonlinear Schr\"odinger (NLS) equation,
generalized Korteweg--de Vries (gKdV), and nonlinear Klein--Gordon (NLKG) arise in
a wide range of physical contexts, including nonlinear optics, plasma physics, and
water waves, see, for example, \cite{Agrawal2001,KA2003,Z1972,ZakRub1973,SS1999,Fibich2015,KS,50years},  
and the references therein.
Of particular interest are {\it solitary waves} or {\it ground states}:
spatially localized solutions that propagate without changing shape, often
organize the long-time dynamics of the flow under consideration, the main components of the {\it soliton resolution} conjecture \cite{Soffer2006,Tao2009}.

In this paper we focus on the {\it focusing}, power-type one-dimensional models:
\begin{align}
    i u_t + u_{xx} + \gamma |u|^{p-1}u &= 0,
    \qquad &&\text{(NLS)}, \label{NLS}\\
    u_t + \big( u_{xx} + \gamma |u|^{p} \big)_x &= 0,
    \qquad &&\text{(gKdV)}, \label{gKdV}\\
    u_{tt} + u_{xx} + b u + \gamma |u|^{p-1} u &= 0,
    \qquad &&\text{(NLKG)}, \label{NLKG}
\end{align}
with parameters $b>0$, $\gamma>0$ and integer nonlinearity exponent $p>1$.
The sign convention in front of the nonlinear term is chosen so that these
equations are focusing (and thus, possess solitary wave or ground state solutions); in contrast, the defocusing case has the opposite sign
in front of the power nonlinearity and does not have 
ground states.

Each of the models \eqref{NLS}--\eqref{NLKG} enjoys conservation laws on its
maximal interval of existence (Cauchy problems are well-understood for these models in the energy space $H^1$, see, for instance, Cazenave~\cite{Cazenave2003}, Tao~\cite{Tao2006}, Linares--Ponce~\cite{LinaresPonce2015}). For NLS \eqref{NLS}, the conserved mass and Hamiltonian (energy) are given by
\[
    M[u(t)] = \int_{\mathbb{R}} |u(t,x)|^2 \, dx,
    \qquad
    E[u(t)] = \int_{\mathbb{R}} \Big( \tfrac{1}{2}|u_x|^2
    - \tfrac{\gamma}{p+1}|u|^{p+1} \Big) dx,
\]
while for gKdV \eqref{gKdV} one has analogous conservation of the $L^2$-norm (often called momentum in that case) and
a suitable Hamiltonian functional.
In the NLKG equation \eqref{NLKG}, the conserved energy in the phase space $(u,u_t)$ is
\[
    E[u(t),u_t(t)] = \int_{\mathbb{R}} \Big(
        \tfrac{1}{2}|u_t|^2 + \tfrac{1}{2}|u_x|^2
        + \tfrac{b}{2}|u|^2 - \tfrac{\gamma}{p+1}|u|^{p+1}
    \Big)\,dx.
\]
Another conserved quantity, often called momentum in NLS and NLKG models and an $L^1$-type integral in the gKdV model, is also available, but not needed in this paper.\footnote{In some specific cases, the NLS and gKdV models are completely integrable, such as 1D cubic NLS or KdV and mKdV, though that is not relevant for this paper.}
\smallskip

Solitary-wave or ground state solutions are obtained by looking for special ansatzes that (typically\footnote{In some higher-dimensional dispersive PDE the soliton profiles can be non-radial, and thus, do not reduce to an ODE problem, for the purpose of this study we only consider radially symmetric profiles, and even further simplification in this paper, the 1D case.}) reduce the PDE to an ODE for a spatial profile $Q$. For the NLS equation \eqref{NLS}, we seek
standing waves of the form
\[
    u(x,t) = e^{i b t} Q(x), \qquad b>0.
\]
Substituting into \eqref{NLS} and using $|e^{ibt}|=1$, we obtain
\begin{equation}\label{E:1}
    Q'' - b Q + \gamma |Q|^{p-1} Q = 0.
\end{equation}
We are interested in $H^1$ ground state solutions to \eqref{E:1}, which are well-known to exist (\cite{Lions84,Wein85modu}), and are unique, positive and radial (symmetric around the origin in 1D). Thus, we study the positive ground-state profiles $Q>0$, hence, we may drop the absolute value and write $|Q|^{p-1}Q = Q^p$ in \eqref{E:1}, which yields
\begin{equation}\label{2ndODE}
    Q'' = b Q - \gamma Q^p.
\end{equation}

For the gKdV equation \eqref{gKdV}, we seek traveling solitary waves
\[
    u(x,t) = Q(x-bt),
\]
with speed $b>0$ (thus, solitary waves travel to the right). Writing $\xi = x-bt$, we compute
\[
    u_t = -b Q'(\xi), \quad
    u_x = Q'(\xi), \quad
    u_{xx} = Q''(\xi), \quad
    |u|^p = |Q(\xi)|^p.
\]
Substituting into \eqref{gKdV} gives
\[
    -b Q'(\xi) + \big(Q''(\xi) + \gamma |Q(\xi)|^p\big)' = 0,
\]
that is,
\[
    (-b Q + Q'' + \gamma |Q|^p)' = 0.
\]
Integrating once in $\xi$ yields
\[
    -b Q + Q'' + \gamma |Q|^p = C,
\]
for some constant $C\in\mathbb{R}$. For localized solitary waves with
$Q(\xi)\to 0$ as $|\xi|\to\infty$, one has $C=0$, and thus,
\[
    Q'' = b Q - \gamma |Q|^p,
\]
which again reduces to \eqref{2ndODE} for positive profiles.

Similarly, for the NLKG equation \eqref{NLKG}, we consider traveling waves of
the form
\[
    u(x,t) = Q(x-ct),
\]
with some speed $c\in\mathbb{R}$. Substituting this ansatz and computing
$u_t,u_{tt},u_{xx}$ leads to
\[
    (1+c^2) Q'' + b Q + \gamma |Q|^p Q = 0.
\]
Dividing by $1+c^2$ and renaming parameters
\[
    \tilde{b} := -\frac{b}{1+c^2}, \qquad
    \tilde{\gamma} := \frac{\gamma}{1+c^2},
\]
one obtains
\[
    Q'' = \tilde{b} \, Q - \tilde{\gamma} \, |Q|^p Q,
\]
which is again of the form \eqref{2ndODE} after dropping absolute values for
positive $Q$ and relabeling $\tilde{b},\tilde{\gamma}$ as $b,\gamma$.

In all three models \eqref{NLS}--\eqref{NLKG}, solitary waves are therefore
described by positive, localized $H^1$ (finite energy) solutions $Q$ of the second-order profile ODE
\eqref{2ndODE}. Under suitable assumptions on $p$, these ground states can be
constructed variationally as minimizers of the energy functional at fixed mass
and are (orbitally) stable in the sense of Weinstein and of Grillakis--Shatah--Strauss; see
Weinstein~\cite{Wein85modu,Wein1986Lya} and
Grillakis--Shatah--Strauss~\cite{GSS87,GSS90}.
For gKdV, well-posedness and scattering, as well as the role of solitary waves,
are discussed in the work of Kenig--Ponce--Vega~\cite{kenig1993gkdv}, see also Bona-Souganidis-Strauss \cite{BSS}.
We emphasize that in the present paper we do not prove new existence or stability
results; instead, we take the profile equation \eqref{2ndODE} as a model for
standing/traveling-wave ODE and focus on its accurate numerical resolution, {\it comparing classical computational methods with the ones that use neural network approaches}.

\subsection{Exact Profiles and Numerical Framework}

We numerically explore the one-dimensional second-order nonlinear stationary-wave ODE
\begin{align}
Q'' &= b Q - \gamma Q^p, \label{2ndODE-again}\\ 
Q'(0) &= 0,\quad Q(\infty)=0, \label{2ndODEBC}
\end{align}
with parameters $b>0$, $\gamma>0$ and integer nonlinearity exponent $p>1$.
The steady-state equation \eqref{2ndODE-again} admits an explicit nontrivial
solution of the form
\begin{equation}\label{Qexact}
Q(x) = A \,\sech^{\frac{2}{p-1}}(\beta x) \equiv Q_{exact}(x),
\end{equation}
where
\begin{align}\label{E:coeff}
A &= A(b,\gamma,p) = 
    \left( \frac{b(p+1)}{2\gamma} \right)^{\frac{1}{p-1}},
\quad \beta = \beta(b,\gamma,p) = \frac{\sqrt{b}(p-1)}{2}.
\end{align}
This explicit profile serves both as a reference solution and as a way to encode
the correct amplitude at the origin. Typically, the parameters $b$ and $\gamma$ are taken to be 1 (or rescaled to 1), however, for the purpose of this paper, we allow some flexibility in these parameters as it can be useful later in training the neural networks.

Stationary solitary-wave profiles have been extensively studied in the context of nonlinear dispersive equations, and a variety of numerical methods have been developed for their computation, including  \cite{AblowitzSegur1981,HH1976, LPSS1987, ADKM1992, Trefethen2000, Boyd2001,LakobaYang2007,Fibich2015}. {\it Classical} numerical methods, such as finite difference (FD) discretization coupled with Newton's iteration or Petviashvili’s  iteration coupled with FFT, provide reliable accuracy and convergence  for solitary-wave solutions of \eqref{2ndODE}, but typically require mesh design and problem-specific tuning. In contrast, {\it neural-network}-based solvers—including the physics-informed neural networks (PINNs) \cite{RPE19}, learn the solution or solution operator directly, enabling mesh-free approximation. The neural operator methods, such as, the Deep Operator Network (DeepONet) \cite{lu2021deeponet} and Fourier Neural Operators (FNOs) \cite{FNO}, offer a very rapid inference across parameter regimes {\it after} the neural network training, at the cost of increased training complexity. For a recent review on machine learning and neural network approaches in nonlinear waves, see \cite{Kev2025} and references therein.
\smallskip

The goal of this paper is to assess the validity of neural-network-based methods (PINN, DeepONet, FNO, etc.) for computing solitary-wave profiles by systematically comparing their accuracy, convergence, and efficiency with established classical approaches. 
Since stationary solitary-wave profiles serve as fundamental building blocks in many nonlinear dispersive equations, we present a systematic comparison between {\it classical} numerical solvers and {\it neural-network}-based approaches on the well-controlled benchmark problem \eqref{2ndODE-again} in this paper. The aim is to assess their respective strengths and limitations. 
\smallskip

Our results show that classical methods compute solitary-wave profiles with high accuracy and efficiency for single-instance problems in 1D, though require mesh generations and repeated nonlinear solves or re-computations of the nonlinear system  if parameters have to be varied or adjusted, which can limit their efficiency in large parametric studies or higher-dimensional extensions. In the one-dimensional setting considered here, classical methods remain the most reliable and accurate approach; PINNs provide a flexible but generally less accurate and more computationally expensive alternative due to costly training and slow convergence. In contrast, operator-learning approaches partially alleviate these limitations by shifting computational cost to an offline training stage and enabling rapid inference for new parameter instances. This comparison clarifies the practical trade-offs between classical and neural approaches in the numerical computation of solitary-wave profiles.
\smallskip

The paper is organized as follows: Section 2 presents the classical numerical methods, along with numerical results, and an exploration of solitary-wave solutions. Section 3 focuses on neural-network-based methods, covering physics-informed neural networks (PINNs) and operator-learning approaches (DeepONet and Fourier Neural Operators), together with a comparison of their performance. Section 4 concludes with a discussion of the main findings. Additional numerical simulation results and extended tables are listed in the Appendix.
\smallskip

{\bf Acknowledgments.} The research for this paper came out of the STEM summer seminar that S.R. and Z.W. organized in Summer 2025, which was partially supported by the NSF grant DUE/EDU-2221491 (PI: S. Roudenko). C.H. and S.R. were also partially supported by the NSF grant DMS-2452782.

\section{Classical methods}
We first compute the numerical solutions of the equations \eqref{2ndODE-again} and \eqref{2ndODEBC} by following classical numerical methods:
\begin{itemize}
\item[1.]  finite-difference with Newton's iteration,
\item[2.] Petviashvili's method.
\end{itemize}
The boundary condition at infinity is approximated at a  large enough number $L$, so  in simulation we solve the following equation  and boundary conditions
\begin{align}
Q'' &= b Q - \gamma Q^p, \label{2ndODE-finite}\\ 
Q'(0) &= 0,\quad Q(L)=0. \label{2ndODEBC-finite}
\end{align}
We choose $L=30$, which is sufficiently large, since the value of the exact solution  $Q(30)$ is on the order of $10^{-13}$, effectively zero for numerical purposes.

The goal of this section is to establish a baseline for numerical performance in terms of accuracy and computational efficiency (measured by CPU time), which is in later section compared against the results obtained from neural-network-based solvers. For the purpose of this paper and a reasonable computational device access,  most of the simulations are performed on a personal computer equipped with an AMD Ryzen 5 7530U processor and 12 GB of RAM\footnote{The purpose is not to use the fastest possible (and costly) computational facilities, but rather make it sufficiently fast and individually accessible.}, we later term it as a `primary' device. In Section \ref{S:secondary} for comparison purposes we also use a `secondary' device, with specifications and comparisons provided there.

\subsection{Set up of finite difference method and the Petviashvili method}
We first test the finite difference method coupled with Newton's iteration for \eqref{2ndODE-finite} and \eqref{2ndODEBC-finite}. Note that the system \eqref{2ndODE-finite}-\eqref{2ndODEBC-finite} admits a trivial solution, in addition to the physically relevant positive nonlinear wave profile.  In order to capture the nonlinear wave, numerically,  one could: 

(1) select an initial guess sufficiently close to the desired solution to ensure Newton’s method converges for mixed boundary condition in \eqref{2ndODE-finite}-\eqref{2ndODEBC-finite};

(2) artificially enforce a Dirichlet boundary condition at 0, i.e.,  use the following Dirichlet boundary conditions (instead of \eqref{2ndODEBC-finite}):
\begin{equation}
    Q(0) = A,\quad Q(L)=0, \label{2ndODEBC-Dirichlet}
\end{equation}
where $A$ is given by the exact solution \eqref{Qexact}-\eqref{E:coeff}. Once the boundary conditions are specified, three-point  or five-point stencil central difference is used to approximate the second derivative (for a desired accuracy). See Algorithm \ref{alg:FD} for the 3-stencil finite difference method.

\begin{center}
\centering
\begin{algorithm}[h] 
\caption{Finite Difference + Newton’s Method for $Q''=bQ-\gamma Q^p$}
\begin{algorithmic}[1] 
\State Discretize computation domain.
\State Three-stencil central differencing with proper boundary conditions
  \[
  F_i(Q) = \frac{Q_{i-1}-2Q_i+Q_{i+1}}{h^2} - bQ_i + \gamma Q_i^p = 0.
  \]
\State Assemble Jacobian $J$ (tri-diagonal).
\State Choose proper initial guess $\mathbf{Q}^{(0)}$. 
\For{$n=0,1,2,\dots$ until convergence}
  \State Compute $F^{(n)}=F(\mathbf{Q}^{(n)})$, $J^{(n)}=J(\mathbf{Q}^{(n)})$.
  \State Solve $J^{(n)} \delta = -F^{(n)}$.
  \State Update $\mathbf{Q}^{(n+1)} = \mathbf{Q}^{(n)} + \delta$.
\EndFor
\State Return $\{Q_i\}_{i=0}^N$.
\end{algorithmic}
\label{alg:FD}
\end{algorithm}
\end{center} 

For $p=3$ (cubic), $b=\gamma=1$, and the initial guess, a perturbation of $Q_{exact}$ from \eqref{Qexact}, i.e.,  
$$
Q_0(x)=k \, Q_{exact}(x),
$$ 
with $k=0.75,\, 0.9,\, 1.1$ for the Newton's iteration,  we show the order of convergence of the finite difference method using either the {\it three-point} stencil or the {\it five-point} stencil in Table \ref{tab:order_FD2} 
with Dirichlet boundary condition \eqref{2ndODEBC-Dirichlet} (left column) or mixed boundary conditions \eqref{2ndODEBC-finite} (middle and right columns). It is observed that the finite difference method {achieves the optimal} order of convergence with the mixed boundary conditions, while only attaining first order convergence for the  Dirichlet boundary conditions due to the $O(h)$ phase shift (see Appendix \ref{App:BC}). 
\begin{table}[h]
\centering
\scriptsize
\setlength{\tabcolsep}{1.3pt}
\renewcommand{\arraystretch}{1.05}

\newcommand{\einf}{\|e\|_{\infty}}
\newcommand{\etwo}{\|e\|_{2}}

\begin{subtable}[t]{0.325\textwidth}
\centering
\resizebox{\linewidth}{!}{%
\begin{tabular}{|c|c|c|c|c|c|}
\hline
$N$ & iters & $\einf$ & $p_{\infty}$ & $\etwo$ & $p_{2}$  \\ \hline
\multicolumn{6}{|c|}{$k=0.75$} \\ \hline
$2^7$ &  8 & 5.399893e-02 & --    & 6.211146e-02 & --     \\
$2^8$ &  9 & 2.532657e-02 & 1.092 & 2.924331e-02 & 1.087  \\
$2^9$ & 10 & 1.231067e-02 & 1.041 & 1.420608e-02 & 1.042  \\
$2^{10}$&11& 6.066360e-03 & 1.021 & 7.003018e-03 & 1.020  \\ \hline
\multicolumn{6}{|c|}{$k=0.9$} \\ \hline
$2^7$ &  8 & 4.255373e-02 & --    & 4.954073e-02 & --     \\
$2^8$ &  9 & 2.261108e-02 & 0.912 & 2.614063e-02 & 0.922  \\
$2^9$ & 10 & 1.163065e-02 & 0.959 & 1.343280e-02 & 0.961  \\
$2^{10}$&11& 5.896548e-03 & 0.980 & 6.809847e-03 & 0.980  \\ \hline
\multicolumn{6}{|c|}{$k=1.1$} \\ \hline
$2^7$ &  9 & 4.255373e-02 & --    & 4.954073e-02 & --     \\
$2^8$ &  9 & 2.261108e-02 & 0.912 & 2.614063e-02 & 0.922  \\
$2^9$ & 10 & 1.163065e-02 & 0.959 & 1.343280e-02 & 0.961  \\
$2^{10}$&11& 5.896548e-03 & 0.980 & 6.809847e-03 & 0.980  \\ \hline
\end{tabular}}
\caption{Dirichlet boundary conditions (\it 3-point  stencil)}
\end{subtable}\hfill
\begin{subtable}[t]{0.325\textwidth}
\centering
\resizebox{\linewidth}{!}{%
\begin{tabular}{|c|c|c|c|c|c|}
\hline
$N$ & iters & $\einf$ & $p_{\infty}$ & $\etwo$ & $p_{2}$  \\ \hline
\multicolumn{6}{|c|}{$k=0.75$} \\ \hline
$2^7$ & 6 & 6.503086e-03 & --    & 7.289336e-03 & --     \\
$2^8$ & 6 & 1.599123e-03 & 2.024 & 1.781979e-03 & 2.032  \\
$2^9$ & 6 & 3.981542e-04 & 2.006 & 4.424091e-04 & 2.010  \\
$2^{10}$&6& 9.945340e-05 & 2.001 & 1.103267e-04 & 2.004  \\ \hline
\multicolumn{6}{|c|}{$k=0.9$} \\ \hline
$2^7$ & 6 & 6.503086e-03 & --    & 7.289336e-03 & --     \\
$2^8$ & 6 & 1.599123e-03 & 2.024 & 1.781979e-03 & 2.032  \\
$2^9$ & 6 & 3.981542e-04 & 2.006 & 4.424091e-04 & 2.010  \\
$2^{10}$&6& 9.945340e-05 & 2.001 & 1.103267e-04 & 2.004  \\ \hline
\multicolumn{6}{|c|}{$k=1.1$} \\ \hline
$2^7$ & 7 & 6.503086e-03 & --    & 7.289336e-03 & --     \\
$2^8$ & 7 & 1.599123e-03 & 2.024 & 1.781979e-03 & 2.032  \\
$2^9$ & 7 & 3.981542e-04 & 2.006 & 4.424091e-04 & 2.010  \\
$2^{10}$&7& 9.945340e-05 & 2.001 & 1.103267e-04 & 2.004  \\ \hline
\end{tabular}}
\caption{Mixed boundary conditions\\ (\it 3-point  stencil)}
\end{subtable}\hfill
\begin{subtable}[t]{0.325\textwidth}
\centering
\resizebox{\linewidth}{!}{%
\begin{tabular}{|c|c|c|c|c|c|}
\hline
$N$ & iters & $\einf$ & $p_{\infty}$ & $\etwo$ & $p_{2}$  \\ \hline
\multicolumn{6}{|c|}{$k=0.75$} \\ \hline
$2^7$ & 18 & 2.541428e-04 & --    & 2.699598e-04 & --     \\
$2^8$ & 18 & 1.631721e-05 & 3.961 & 1.699436e-05 & 3.990  \\
$2^9$ & 18 & 1.027240e-06 & 3.990 & 1.058877e-06 & 4.004  \\
$2^{10}$&18& 6.432078e-08 & 3.997 & 6.594289e-08 & 4.005  \\ \hline
\multicolumn{6}{|c|}{$k=0.9$} \\ \hline
$2^7$ & 17 & 2.541428e-04 & --    & 2.699598e-04 & --     \\
$2^8$ & 17 & 1.631721e-05 & 3.961 & 1.699436e-05 & 3.990  \\
$2^9$ & 17 & 1.027240e-06 & 3.990 & 1.058877e-06 & 4.004  \\
$2^{10}$&17& 6.432075e-08 & 3.997 & 6.594298e-08 & 4.005  \\ \hline
\multicolumn{6}{|c|}{$k=1.1$} \\ \hline
$2^7$ & 20 & 2.541428e-04 & --    & 2.699598e-04 & --     \\
$2^8$ & 20 & 1.631721e-05 & 3.961 & 1.699436e-05 & 3.990  \\
$2^9$ & 20 & 1.027240e-06 & 3.990 & 1.058877e-06 & 4.004  \\
$2^{10}$&20& 6.432084e-08 & 3.997 & 6.594303e-08 & 4.005  \\ \hline
\end{tabular}}
\caption{Mixed boundary conditions\\ (\it 5-point  stencil)}
\end{subtable}
\caption{\small Comparison of the convergence order for the Dirichlet type and mixed boundary conditions for the {\it 3-point} central stencil FD-Newton scheme (A) and (B) and {\it 5-point} central stencil FD-Newton scheme (C) vs. grid number $N_{interior}$ for \eqref{2ndODE-finite} with variable ($k$) amplitude initialization for $p=3$ (cubic), $Q_0=k\,Q_{exact}(x)$ with $k=0.75,\,0.9,\,1.1$, $L=30,$ and $Tol=10^{-12}$. Here, $\einf$ and $p_\infty$ denote the error and the convergence order in the $L^\infty$ norm, while $\etwo$ and $p_2$ denote the corresponding quantities in the $L^2$ norm.
}
\label{tab:order_FD2}
\end{table}

\begin{Remark}\label{sec:FD}
While the mixed boundary conditions \eqref{2ndODEBC-finite} offer better accuracy, the Dirichlet boundary conditions are easier to implement, especially, for { initializations in} the neural networks in PINN, DeepONet, etc. Since our main concern is to validate the neural network solutions by the classical numerical ones, we mainly explore the classical solution using the Dirichlet boundary conditions. 
\end{Remark}

\subsection{Comparison of finite difference method and the Petviashvili method}
We also consider the classical Petviashvili method for solitary wave computation \cite{YangLakoba2007}, see Algorithm \ref{alg:Petviashvili} for the method utilizing the Fourier transform.

\begin{center}
\centering
\begin{algorithm}[h] 
\caption{Petviashvili’s Method for $Q''=bQ-\gamma Q^p$}
\begin{algorithmic}[1] 
\State Define linear operator $LQ = Q'' -b Q$, nonlinear term $N(Q) = -\gamma Q^p$.
\State Choose proper initial guess $\mathbf{Q}^{(0)}$.
\For{$n=0,1,2,\dots$ until convergence}
  \State Compute nonlinear term $N(Q^{(n)})$.
  \State Compute stabilization factor
    \[
    S^{(n)} = \frac{\langle LQ^{(n)}, Q^{(n)} \rangle}
                   {\langle N(Q^{(n)}), Q^{(n)} \rangle}.
    \]
  \State Update solution in Fourier (or discretized) space:
    \[
    Q^{(n+1)} = \left(L^{-1} N(Q^{(nk)})\right)\,\big(S^{(n)}\big)^{{\Gamma}},
    \]
         with ${\Gamma}=p/(p-1)$.
\EndFor
\State Return $Q(x)$.
\end{algorithmic}
\label{alg:Petviashvili}
\end{algorithm}
\end{center}

It is well-known that the Petviashivili's method converges to a stabilized solution exponentially in iteration when initialized with an appropriate solitary-wave profile \cite{YangLakoba2007}. We now compare the performance of finite difference method and the Petviashvili method by testing \eqref{2ndODE-finite} with $\gamma=1$, $p=3$ (cubic nonlinearity), and $b=1$ on finite interval $[0,L]$. In our simulation we use the Dirichlet boundary conditions 
\begin{align}\label{BC:1}
Q(0)&= A,\quad Q(L)=0,
 \end{align}
and set $L=30$ as the far field. Unless otherwise specified, we  use a uniform partition of $N$ grids, and a maximum iteration of $200$.  Note that  a symmetric  domain $[-L,L]$ is used for the Petviashvili’s method, so that the Fast Fourier Transform (FFT) could be utilized. This approach also helps avoid certain numerical stability issues that arise when Dirichlet boundary conditions are imposed on asymmetric finite domains in solitary-wave computations; see, e.g., \cite{PelinovskyYang2005,drazin1989solitons,yang2010nonlinear}.

\begin{figure}[h!]
    \centering
    \includegraphics[width=0.49\textwidth]{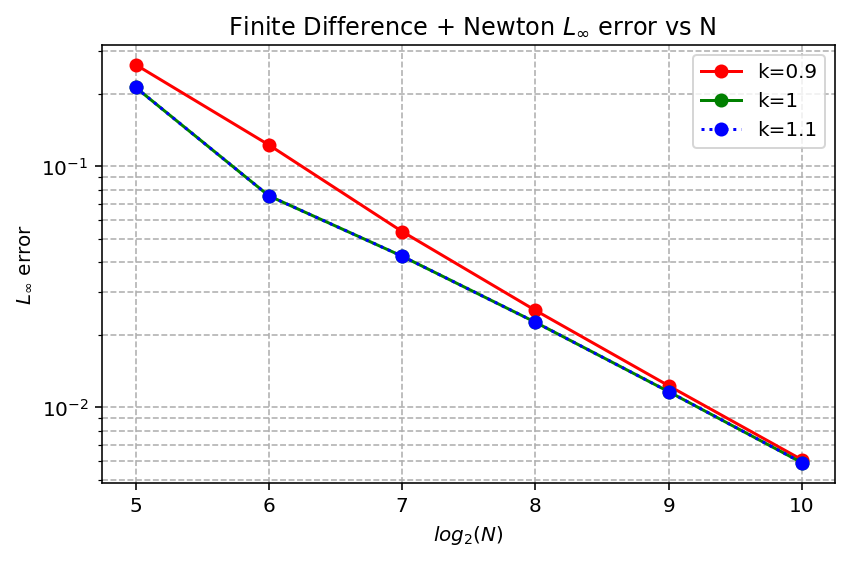}
       \includegraphics[width=0.49\textwidth]{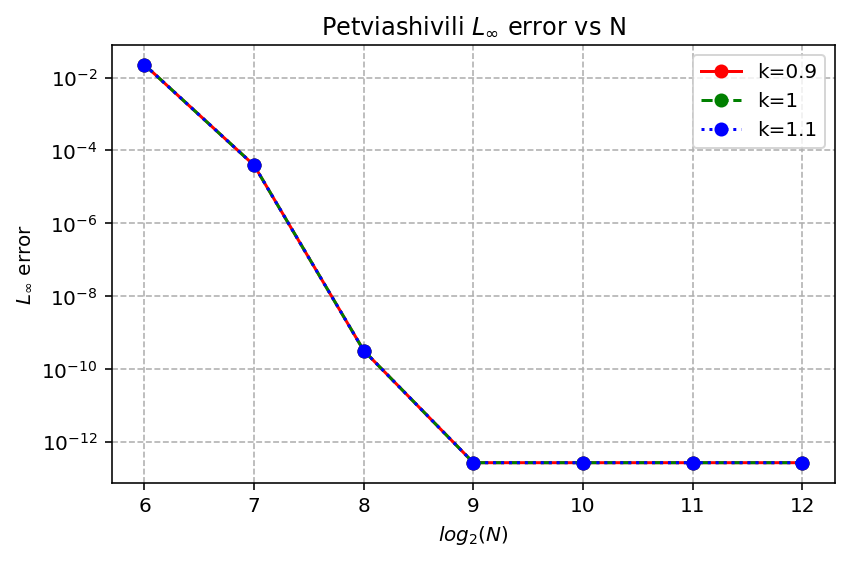}
    \caption{ $L_\infty$ error of the FD-Newton scheme (left) and   the Petviashvili method (right) vs. grid number $N_{interior}$ (in log scale) for \eqref{2ndODE-finite} with variable amplitude initialization, $Q_0 = kQ_{exact}(x),\,  k=0.9,\, 1.0,\, 1.1$, $Tol=10^{-12}$.}
    \label{fig:erroranalysis_FD_Pet_N}
\end{figure}

Figure \ref{fig:erroranalysis_FD_Pet_N}  shows the $L^{\infty}$ error versus the grid $N$ (in log scale) for the finite difference method  (left plot) and the Petviashvili method (right plot), respectively.  We observe that for different initial guesses of the form $Q^{(0)}(x)=k\, Q_{exact}(x)$ with $k=0.9,\,1.0,\, 1.1$, the  $L^{\infty}$ norm decreases as we refine the mesh for both methods.  Note that the error of the Petviashvili method levels out at $N_{interior}=2^9$ due to its exponential error decay, which achieves the order of the machine error for $N\geq 2^9$.

\begin{figure}[h!]
    \centering
    \includegraphics[width=0.49\textwidth]{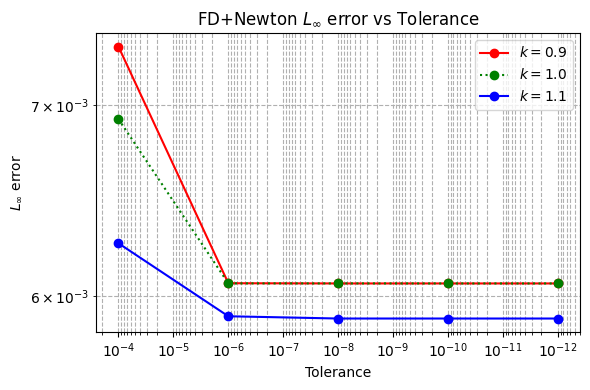}
       \includegraphics[width=0.49\textwidth]{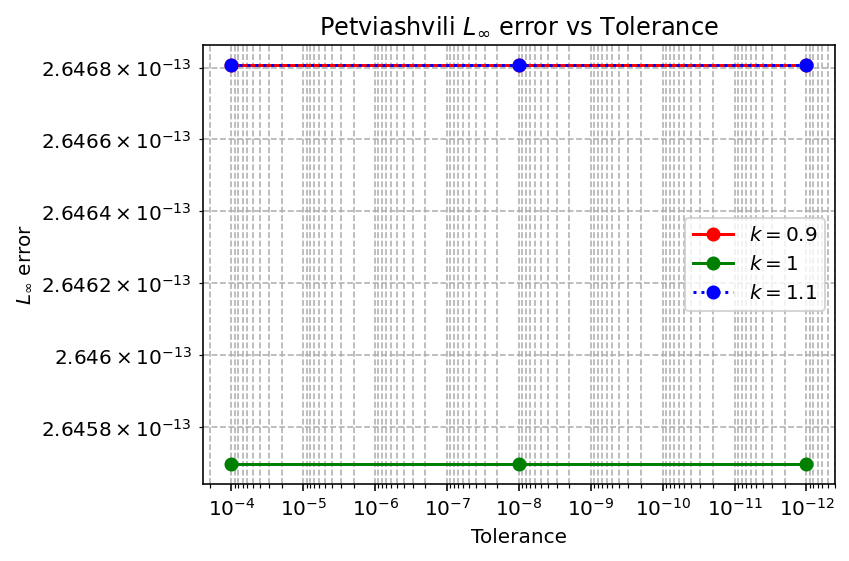}
    \caption{ $L_\infty$ error of the FD-Newton scheme (left) and   the Petviashvili method (right) vs. tolerance $Tol$ for \eqref{2ndODE-finite} with variable amplitude initialization, $Q_0 = kQ_{exact}(x),\,  k=0.9,\, 1.0,\, 1.1$.}
    \label{fig:tol}
\end{figure}

We further investigate the effect of the tolerance $Tol$ in  Figure \ref{fig:tol}, which shows the effect of it on the $L^{\infty}$ error. Note that a sufficiently small $Tol$ ($<10^{-7}$) guarantees the optimal accuracy for finite difference method with $N=2^9$.  We also observe that a tolerance $Tol=10^{-4}$ is sufficient for the Petviashvili method. 

\begin{figure}[h]
    \centering
    \includegraphics[width=0.49\textwidth]{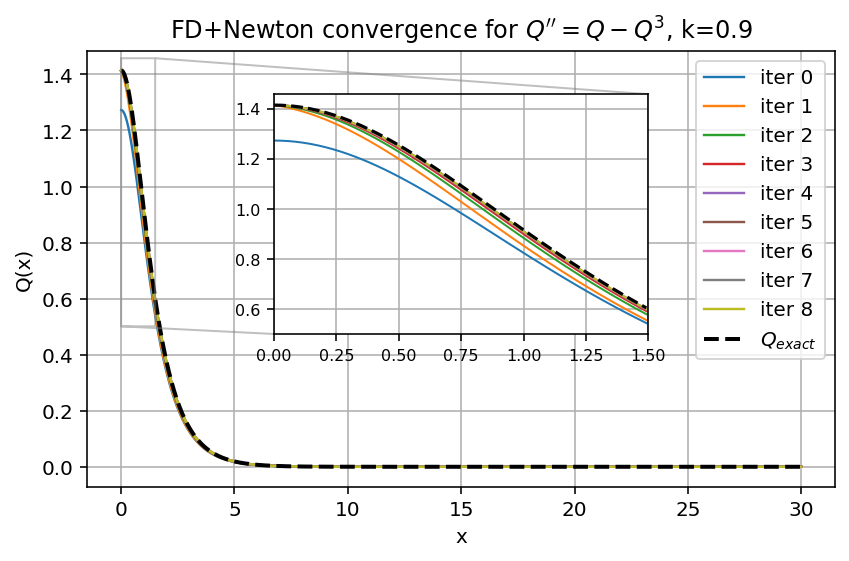}
    \includegraphics[width=0.49\textwidth]{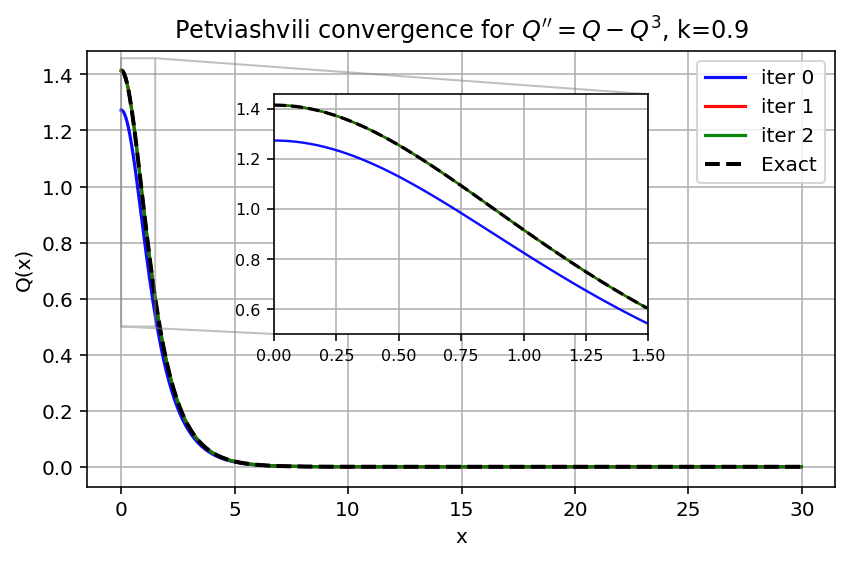}
    \caption{Comparison of convergences with iterations for the FD+Newton (left) and Petviashvili's (right) methods for \eqref{2ndODE-finite} with $p=3$ and 
    $u_0 = 0.9Q_{exact}(x)$.}
    \label{fig:erroranalysis_FD_Pet_iter}
\end{figure}

Figure \ref{fig:erroranalysis_FD_Pet_iter} 
shows an approximate solution for $Q(x)$ after each iteration. With $Tol=10^{-12}$, the finite difference method and Petviashvili method converge after $8$ and $2$ iterations, respectively.   For consistency and accuracy, in our numerical convergence experiments we, therefore, select a tolerance of $Tol=10^{-12}$, unless otherwise specified. 

\subsection{Exploration of the ODE with classical methods} After confirming the convergence of both the finite difference and Petviashvili's methods, we next explore the behavior of solutions to 
\begin{equation}
    Q''=bQ-\gamma Q^p \label{2ndODEagain}
\end{equation}
in various settings. Since the solution of \eqref{2ndODEagain} is mainly determined by the power $p$, we set $b=\gamma=1$ and vary the power $p$.  We test how the choice of the initial guess $Q^{(0)}$ in {\bf Algorithms \ref{alg:FD}} and {\bf\ref{alg:Petviashvili}} affects the stability, accuracy and efficiency of these two schemes. For that we consider 4 types of initial conditions $U_0^{(\cdot)}$, which are physically-motivated and described in Table \ref{initials}: power of a sech function, Gaussian, super-Gaussian (these three are smooth and exponentially decreasing) and a Hat function, a linear function decreasing from $A$ down to $0$ on the interval $[0,L]$ (thus, continuous but not differentiable example of an initial guess). We test both methods with these data perturbing either the amplitude $A$ (thus, denoted by $U^{(A)}_0$) or the effective width (denoted by $U^{(W)}_0$) or the power (denoted by $U^{(P)}_0$) by a perturbation parameter $k$, for which we take the following variations: $k=0.5, 0.75,0.9, 1, 1.1, 1.25$. The complete results are given in Appendix in Tables \ref{tab:appendix_1}-\ref{tab:appendix_end}, below we give a concise summary of them. 
{\small
\begin{table}[h!]
\centering 
\renewcommand{\arraystretch}{1.3}
\begin{tabular}{|l|l|l|l|} \hline 
\textbf{Type} & \textbf{Variable Amplitude} & \textbf{Variable Width}
& \textbf{Variable Power} \\ \hline
Sech/Soliton 
& \( U_0^{(A)} = kA\, \sech^\frac{2}{p-1}\!\left({\tfrac{p-1}{2}x}\right) \)
& \( U_0^{(W)} = A\, \sech^\frac{2}{p-1}\!\left({\tfrac{p-1}{2}kx}\right) \) 
& \( U_0^{(P)} = A\, \sech^\frac{2k}{p-1}\!\left({\tfrac{p-1}{2}x}\right) \) \\ 
\hline
Gaussian 
& \( U^{(A)}_0 = kA\, e^{-x^{2}/A} \)
& \( U^{(W)}_0 = A\, e^{-k\,x^{2}/A} \) 
& $N.A.$ \\  \hline
Super-Gaussian 
& \( U^{(A)}_0 = kA\, e^{-x^{4}/A} \)
& \( U^{(W)}_0 = A\, e^{-k\,x^{4}/A} \) 
& $N.A.$\\ \hline
Hat Function 
& \(U_0^{(A)}=kA\chi_{|x|\leq L}\left(1-\frac{x}{L}\right)\)
& \(U_0^{(W)}=kA\chi_{|x|\leq \frac{L}{k}}\left(1-\frac{xk}{L}\right)\)
& $N.A.$\\ \hline
\end{tabular}
\caption{Types of initial guesses used in both finite difference and Petviashvili methods, with $A$ as in \eqref{Qexact} and variable $k$.}
\label{initials}
\end{table}
}

We summarize our results in Tables \ref{tab:NP_p_tenninth} - \ref{tab:NP_k1p1p3}, where we included the number of iterations, CPU time, $L^{\infty}$ and $L^2$ errors.  We test both the finite difference and Petviashvili’s methods for different types of the initial guesses $U_0$ (from Table \ref{initials}) 
and varying nonlinearity $p$ or the parameter $k$. 
Note that we did not use the Hat initial guess for the finite difference method, as its second derivative vanishes, and thus, not applicable.  
\begin{table}[ht!]
\centering
\scriptsize
\setlength{\tabcolsep}{1.95pt}
\renewcommand{\arraystretch}{1.05}
\begin{subtable}[t]{0.29\textwidth}
\centering
\begin{tabular}{|c|}
\hline
Initialization \\ \hline\hline
Sech Variable Amplitude\\\hline
Sech Variable Width\\\hline
Sech Variable Power \\\hline 
Gaussian Variable Amplitude\\\hline
Gaussian Variable Width\\\hline
Super Gaussian Variable Amplitude\\\hline
Super Gaussian Variable Width\\\hline
Hat Function Variable Amplitude\\\hline
Hat Function Variable Width\\\hline
\end{tabular}
\end{subtable}
\hfill
\begin{subtable}[t]{0.36\textwidth}
\centering
\begin{tabular}{|c|c|c|c|}
\hline
iter & CPU time & $L^\infty$ error & $L^2$ error \\ \hline\hline
9 & 0.406250 & 1.933107e-03 & 4.385094e-03\\ \hline
10 & 0.484375 & 1.933107e-03 & 4.385094e-03\\ \hline
9 & 0.453125 & 1.933106e-03 & 4.385094e-03\\ \hline
14 & 0.687500 & 1.943589e-03 & 4.408515e-03\\ \hline
14 & 0.671875 & 1.943589e-03 & 4.408515e-03\\ \hline
14 & 0.671875 & 1.943589e-03 & 4.408515e-03\\ \hline
14 & 0.671875 & 1.943589e-03 & 4.408515e-03\\ \hline
$ -$ &$ -$ & $ -$ & $ -$ \\ \hline
$- $ & $ -$ & $ -$ & $ -$ \\ \hline

\end{tabular}
\caption{FD + Newton}
\end{subtable}
\hfill
\begin{subtable}[t]{0.32\textwidth}
\centering
\begin{tabular}{|c|c|c|c|}
\hline
iter & CPU time& $L^\infty$ error & $L^2$ error \\ \hline\hline
6 & $<10^{-8}$ & 2.169142e-08 & 2.278022e-08\\ \hline
232 & 0.062500 & 2.169161e-08 & 2.278066e-08\\ \hline
275 & 0.062500 & 2.169161e-08 & 2.278066e-08\\ \hline
247 & 0.046875 & 2.169161e-08 & 2.278066e-08\\ \hline
247 & 0.046875 & 2.169161e-08 & 2.278066e-08\\ \hline
247 & 0.093750 & 2.169161e-08 & 2.277864e-08\\ \hline
247 & 0.093750 & 2.169161e-08 & 2.277864e-08\\ \hline
265 & 0.062500 & 2.169161e-08 & 2.278066e-08\\ \hline
267 & 0.062500 & 2.169161e-08 & 2.278066e-08\\ \hline
 
\end{tabular}
\caption{Petviashvili, factor $\Gamma=10$}
\end{subtable}
\caption{Finite difference vs. Petviashvili method:  comparison of initial conditions  from Table \ref{initials} with $p=\frac{10}9$, $k=0.9$, $N=2^{10}$, $L=30$, $Tol=10^{-12}$ for $Q'' = Q - Q^{10/9}$.}
\label{tab:NP_p_tenninth}
\end{table}

In Tables \ref{tab:NP_p_tenninth} - \ref{tab:NP_k1p1p3}, we test the effect of nonlinearity $p$ in the equation \eqref{2ndODEagain} with different types of initial guesses $U_0(x)$ from Table \ref{initials} while fixing $k=0.9$. Specifically, 
$p=10/9$ is in Table \ref{tab:NP_p_tenninth}, $p=7$  is in Table \ref{tab:NP_p7}) and $p=3$ is in table Table \ref{tab:NP_k1p1p3} (there we also compare changing parameter $k$).

Observe that the finite difference coupled with Newton's method converges with similar CPU times and number of iterations, resulting in similar numerical errors (on the order of $10^{-3}$) for the same given convergence criterion. {This shows that the method is insensitive to the power $p$, which is suitable if one needs to consider small or large values of power nonlinearity $p$.}  We note that the finite difference method does not converge if the Hat function initializations are used, which is due to the fact that the central difference vanishes for a linear function. The Petviashvili method generally outperforms the finite difference method in terms of the CPU time and accuracy (machine precision, around $10^{-13}$), and it converges when non-smooth Hat function initial guess is used. It is, however, interesting to observe that the number of iterations in the Petviashvili method depends on the power $p$: the smaller $p$ values require significantly more iterations, which can be explained due to their slower decay (compare the powers of sech in dependence of $p$) than the larger $p$ values.

\begin{table}[h]
\centering
\scriptsize
\setlength{\tabcolsep}{1.95pt}
\renewcommand{\arraystretch}{1.05}
\begin{subtable}[t]{0.29\textwidth}
\centering
\begin{tabular}{|c|}
\hline
Initialization \\ \hline\hline
Sech Variable Amplitude\\\hline
Sech Variable Width\\\hline
Sech Variable Power \\\hline 
Gaussian Variable Amplitude\\\hline
Gaussian Variable Width\\\hline
Super Gaussian Variable Amplitude\\\hline
Super Gaussian Variable Width\\\hline
Hat Function Variable Amplitude\\\hline
Hat Function Variable Width\\\hline
\end{tabular}
\end{subtable}
\hfill
\begin{subtable}[t]{0.36\textwidth}
\centering
\begin{tabular}{|c|c|c|c|}
\hline
iter & CPU time & $L^\infty$ error & $L^2$ error \\ \hline\hline
8 & 0.375000 & 7.560160e-03 & 7.129116e-03\\ \hline
7 & 0.343750 & 7.086748e-03 & 6.688925e-03\\ \hline
6 & 0.296875 & 7.086748e-03 & 6.688925e-03\\ \hline
8 & 0.390625 & 7.560160e-03 & 7.129116e-03\\ \hline
8 & 0.375000 & 7.086748e-03 & 6.688925e-03\\ \hline
9 & 0.437500 & 7.086748e-03 & 6.688925e-03\\ \hline
10 & 0.484375 & 7.086748e-03 & 6.688925e-03\\ \hline
$ -$ &$ -$ & $ -$ & $ -$ \\ \hline
$- $ & $ -$ & $ -$ & $ -$ \\ \hline
\end{tabular}
\caption{FD + Newton}
\end{subtable}
\hfill
\begin{subtable}[t]{0.32\textwidth}
\centering
\begin{tabular}{|c|c|c|c|}
\hline
iter & CPU time& $L^\infty$ error & $L^2$ error \\ \hline\hline
2 & $<10^{-8}$ & 1.485527e-13 & 1.484803e-13\\ \hline
27 & 0.015625 & 5.967449e-13 & 8.196293e-13\\ \hline
29 & 0.015625 & 3.952394e-13 & 5.536132e-13\\ \hline
28 & 0.015625 & 4.328760e-13 & 6.030365e-13\\ \hline
28 & 0.015625 & 5.762057e-13 & 7.922620e-13\\ \hline
29 & 0.015625 & 6.493694e-13 & 8.893390e-13\\ \hline
30 & 0.015625 & 2.775558e-13 & 4.025816e-13\\ \hline
32 & $<10^{-8}$ & 3.625988e-13 & 5.112039e-13\\ \hline
32 & 0.015625 & 3.900213e-13 & 5.463048e-13\\ \hline
\end{tabular}
\caption{Petviashvili, factor $\Gamma=\frac{7}{6}$}
\end{subtable}
\caption{Finite difference vs. Petviashvili method:  comparison of initial conditions  from Table \ref{initials} with $p=7$, $k=0.9$, $N=2^{10}$, $L=30$, $Tol =10^{-12}$ for $Q'' = Q - Q^7$.}
\label{tab:NP_p7}
\end{table}
\begin{table}[h]
\centering
\scriptsize
\setlength{\tabcolsep}{1.95pt}
\renewcommand{\arraystretch}{1.05}
\begin{subtable}[t]{0.29\textwidth}
\centering
\begin{tabular}{|c|}
\hline
Initialization \\ \hline\hline
Sech Variable Amplitude\\\hline
Sech Variable Width\\\hline
Sech Variable Power \\\hline 
Gaussian Variable Amplitude\\\hline
Gaussian Variable Width\\\hline
Super Gaussian Variable Amplitude\\\hline
Super Gaussian Variable Width\\\hline
Hat Function Variable Amplitude\\\hline
Hat Function Variable Width\\\hline
\end{tabular}
\end{subtable}
\hfill
\begin{subtable}[t]{0.36\textwidth}
\centering
\begin{tabular}{|c|c|c|c|}
\hline
iter & CPU time & $L^\infty$ error & $L^2$ error \\ \hline\hline
 8  & 0.390625 & 6.060285e-03 & 6.996089e-03 \\ \hline
 8  & 0.375000 & 5.890892e-03 & 6.803295e-03 \\ \hline
 7  & 0.343750 & 4.834268e-03 & 6.486603e-03 \\ \hline
 9  & 0.437500 & 6.060285e-03 & 6.996089e-03 \\ \hline
 9  & 0.453125 & 6.060285e-03 & 6.996089e-03 \\ \hline
10  & 0.468750 & 6.060285e-03 & 6.996089e-03 \\ \hline
11  & 0.531250 & 6.060285e-03 & 6.996089e-03 \\ \hline
$ -$ &$ -$ & $ -$ & $ -$ \\ \hline
$- $ & $ -$ & $ -$ & $ -$ \\ \hline
\end{tabular}
\caption{FD + Newton}
\end{subtable}
\hfill
\begin{subtable}[t]{0.32\textwidth}
\centering
\begin{tabular}{|c|c|c|c|}
\hline
iter & CPU time& $L^\infty$ error & $L^2$ error \\ \hline\hline
2 & $<10^{-8}$ & 2.646808e-13 & 2.646294e-13 \\ \hline
36 & 0.015625 & 9.244827e-13 & 1.571286e-12\\ \hline
36 & 0.015625 & 5.442313e-13 & 9.486899e-13\\ \hline
38 & 0.015625 & 5.110357e-13 & 8.952522e-13\\ \hline
37 & 0.031250 & 8.508749e-13 & 1.450222e-12\\ \hline
37 & 0.015625 & 7.617240e-13 & 1.303072e-12\\ \hline
37 & 0.015625 & 6.602496e-13 & 1.137017e-12\\ \hline
41 & $<10^{-8}$ & 9.958701e-13 & 1.689112e-12\\ \hline
43 & 0.015625 & 5.329071e-13 & 9.293605e-13\\ \hline
\end{tabular}
\caption{Petviashvili, factor $\Gamma=\frac{3}{2}$}
\end{subtable}
\scriptsize
\setlength{\tabcolsep}{1.95pt}
\renewcommand{\arraystretch}{1.05}
\begin{subtable}[t]{0.29\textwidth}
\centering
\begin{tabular}{|c|}
\hline
Initialization \\ \hline\hline
Sech Variable Amplitude\\\hline
Sech Variable Width\\\hline
Sech Variable Power \\\hline 
Gaussian Variable Amplitude\\\hline
Gaussian Variable Width\\\hline
Super Gaussian Variable Amplitude\\\hline
Super Gaussian Variable Width\\\hline
Hat Function Variable Amplitude\\\hline
Hat Function Variable Width\\\hline
\end{tabular}
\end{subtable}
\hfill
\begin{subtable}[t]{0.36\textwidth}
\centering
\begin{tabular}{|c|c|c|c|}
\hline
iter & CPU time & $L^\infty$ error & $L^2$ error \\ \hline\hline
 8 & 0.359375 & 6.139193e-03 & 7.069814e-03\\ \hline 
 8 & 0.359375 & 6.139193e-03 & 7.069814e-03\\ \hline 
 8 & 0.375000 & 6.139193e-03 & 7.069814e-03\\ \hline 
 9 & 0.406250 & 6.139193e-03 & 7.069814e-03\\ \hline 
 9 & 0.421875 & 6.139193e-03 & 7.069814e-03\\ \hline 
10 & 0.468750 & 6.060285e-03 & 6.996089e-03\\ \hline 
10 & 0.484375 & 6.060285e-03 & 6.996089e-03\\ \hline 
$ -$ &$ -$ & $ -$ & $ -$ \\ \hline
$- $ & $ -$ & $ -$ & $ -$ \\ \hline
\end{tabular}
\caption{FD + Newton}
\end{subtable}
\hfill
\begin{subtable}[t]{0.32\textwidth}
\centering
\begin{tabular}{|c|c|c|c|}
\hline
iter & CPU time& $L^\infty$ error & $L^2$ error \\ \hline\hline
33 & 0.015625 & 4.406475e-13 & 7.716050e-13\\ \hline 
33 & 0.015625 & 4.406475e-13 & 7.716050e-13\\ \hline 
33 & 0.015625 & 4.406475e-13 & 7.716050e-13\\ \hline 
34 & 0.015625 & 4.406475e-13 & 7.716050e-13\\ \hline 
34 & 0.015625 & 4.406475e-13 & 7.716050e-13\\ \hline 
37 & 0.015625 & 8.499867e-13 & 1.448754e-12\\ \hline 
37 & 0.015625 & 8.499867e-13 & 1.448754e-12\\ \hline 
42 & 0.015625 & 9.692247e-13 & 1.644796e-12\\ \hline 
42 & 0.015625 & 9.692247e-13 & 1.644796e-12\\ \hline 
\end{tabular}
\caption{Petviashvili, factor $\Gamma=\frac{3}{2}$}
\end{subtable}
\caption{Finite difference vs. Petviashvili method:  comparison of initial conditions  from Table \ref{initials} with $p=3$, $k=0.9$ (top) and $k=1.1$ (bottom), $N=2^{10}$, $L=30$, $Tol=10^{-12}$ for $Q'' = Q - Q^3$.}
\label{tab:NP_k1p1p3}
\end{table}
In Table 
\ref{tab:NP_k1p1p3}, we fix $p=3$ and test the effect of the scaling parameter $k$ for different types of initial guesses $U_0(x)$ from Table \ref{initials}. We observe that the convergence of both methods are insensitive to the scaling parameter $k$, i.e., as long as we start with a reasonable initial guess, both methods converge with the expected accuracy. Once again the finite difference method fails to converge for the Hat function initials due to the vanishing of central difference for linear functions. 
\smallskip

We mention that more details from numerical experiments on these two classical methods are listed in Tables \ref{tab:appendix_1}-\ref{tab:appendix_end} in Appendix \ref{App:2} and \ref{App:3}. 
\smallskip

Having described the details of the classical numerical approaches in the context of \eqref{2ndODE}, we can now compare them with the neural network approximations.

\section{Neural network approximations}

Overall the finite difference methods combined with Newton or Petviashvili iterations can achieve high accuracy for stationary profile ODEs, when some proper initial guesses are provided. However, they both typically require repeated nonlinear solvers as parameters vary, which can be computationally demanding. Extending these methods to higher spatial dimensions or more complex geometries further increases computational cost and implementation complexity due to mesh construction and nonlinear solver requirements. Neural-network-based solvers, such as PINNs provide a flexible, mesh-free framework for approximating solutions of differential equations by enforcing the governing equations and boundary conditions through the training loss. Such methods can be extended to higher dimensions in principle, though training complexity increases substantially, particularly, for problems with localized structures.

In the previous section we use classical methods to capture the profiles of the solitary wave or ground state solutions of \eqref{2ndODE}. In this section to obtain those profiles we explore and compare different types the neural-network based methods.

\subsection{Physics-Informed Neural Network (PINN)} In the PINN framework \cite{RPE19}, we approximate the solution $Q(x)$ by a  standard feedforward multilayer perceptron (MLP) $Q_{\theta}(x)$ of $m$ hidden layers with $n$ neurons in each layer, i.e.,
\begin{equation}
    Q(x) \approx Q_\theta(x),
\end{equation}
where $\theta$ denotes all trainable parameters. The goal of the PINN is to train the neural network $Q_{\theta}(x)$ that  minimizes the residual, based on the equation \eqref{2ndODE-finite}, namely,
\begin{equation}
    \mathcal{R}(x;\theta) = Q''_\theta(x) - b Q_\theta(x) + \gamma Q_\theta^p(x),
    \label{eq:residual}
\end{equation}
while enforcing the boundary conditions \eqref{2ndODEBC-finite}.  With the total loss function defined as 
\begin{equation}
    \mathcal{L}(\theta) =  \frac{1}{N_{cb}}\sum_{i=1}^{N_f} 
    \left| \mathcal{R}(x_i;\theta) \right|^2 +
        \left| Q'_\theta(0) \right|^2 + 
        \left| Q_\theta(L) \right|^2,
\end{equation}
the neural network parameters $\theta$ are optimized by minimizing the total loss:
\begin{equation}
    \theta^* = \arg\min_\theta \mathcal{L}(\theta).
\end{equation}
Table \ref{tab:pinn_hyperparams} lists the setting of the neural network used in the simulations.
{
\begin{table}[h]
\centering
\begin{tabular}{ll}
\toprule
\textbf{Parameter} & \textbf{Value / Description} \\
\midrule
Domain length $L$ & $30$ \\
Number of hidden layers & $3-5$ \\
Neurons per layer & $32$ or $64$ \\
Activation function & $\tanh(x)$ or $\mathrm{SiLU}$ \\
Number of collocation points $N_f$ & $2^7-2^{12}$ \\
Boundary points $N_b$ & $2$ \\
Optimizer & Adam  \\
Learning rate ($lr$) & $10^{-3}$  \\
Epochs & up to $30{,}000$ \\
\bottomrule
\end{tabular} 
\caption{ PINN architecture and  hyperparameters for solving \eqref{2ndODE-finite}-\eqref{2ndODEBC-finite} on $[0,30]$.}
\label{tab:pinn_hyperparams}
\end{table}
}

\subsubsection{Cubic nonlinearity via PINNs}

We start to investigate the validity of solutions obtained via PINN (refer as PINN solution) with $b=1$, $\gamma=1$ and $p=3$ (cubic) 
following the settings given in Table \ref{tab:pinn_hyperparams} and first choose  $\tanh(x)$ as an activation function.
\smallskip

\begin{figure}[h]
    \centering
    \includegraphics[width=0.49\textwidth,height=.33\textwidth]{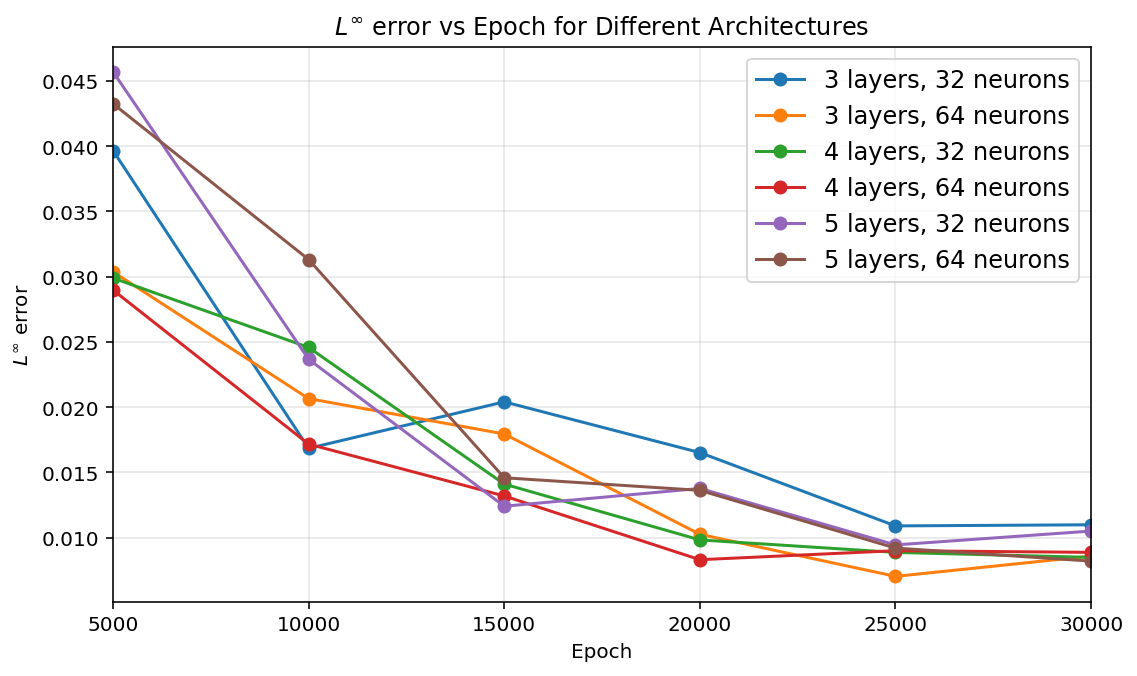}
       \includegraphics[width=0.49\textwidth,height=.33\textwidth]{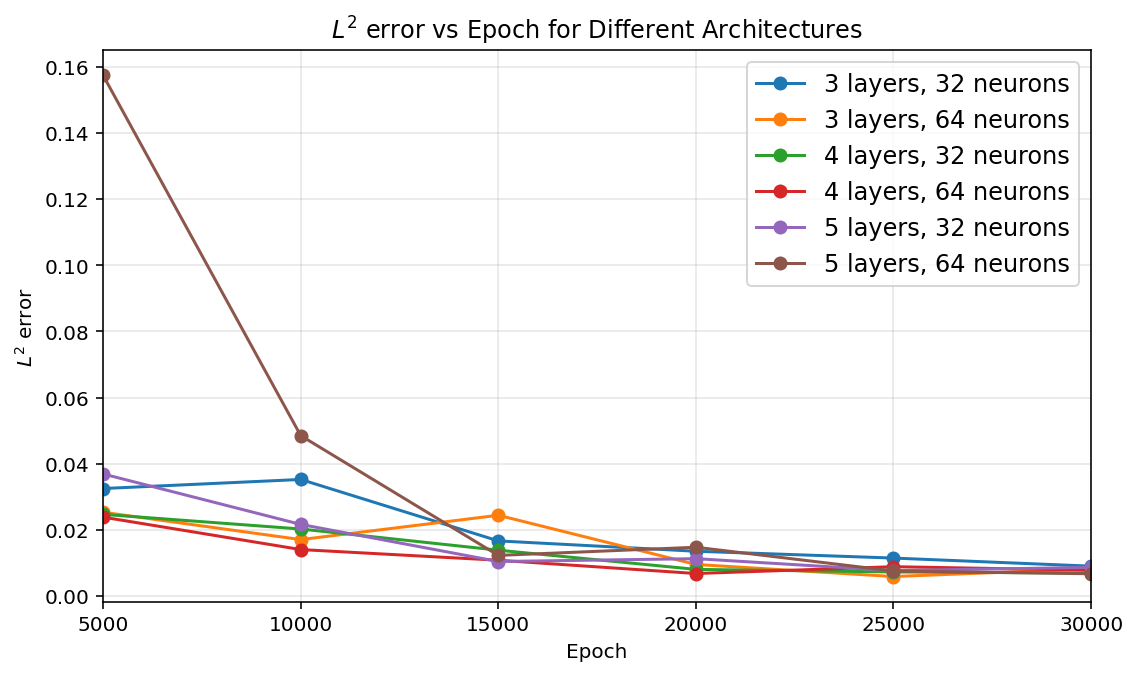}
    \caption{ $L^\infty$ error (left) and $L^2$ error (right) of the PINN with $\tanh$ activation vs. number of epochs for \eqref{2ndODE-finite} for different architecture combinations of layers and neurons.}
    \label{fig:erroranalysis_PINN}
\end{figure}

Figure \ref{fig:erroranalysis_PINN} shows the convergence of the PINN solution with various number of layers $m$ and  number of neurons per layer $n$. One can observe that both the $L^\infty$ and $L^2$ errors decrease when either $m$ or $n$ is increased. The neural network is stable (i.e., the loss function decreases monotonically with number of epochs) in training up to at least $30,000$ epochs. 

\begin{figure}[h]
    \centering   
    \includegraphics[width=0.49\textwidth,height=.3\textwidth]{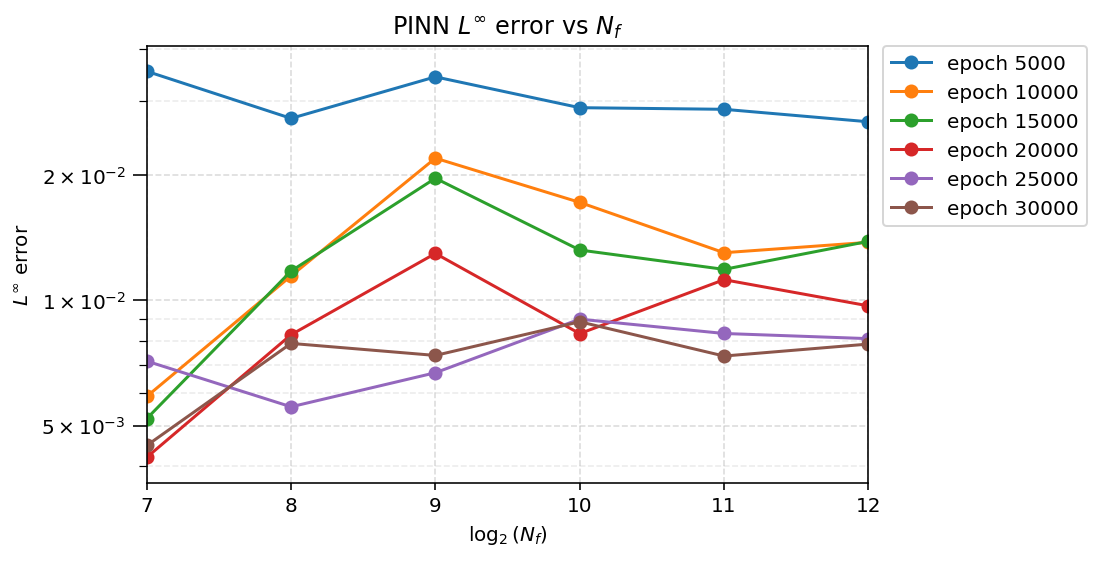}
    \includegraphics[width=0.49\textwidth,height=.3\textwidth]{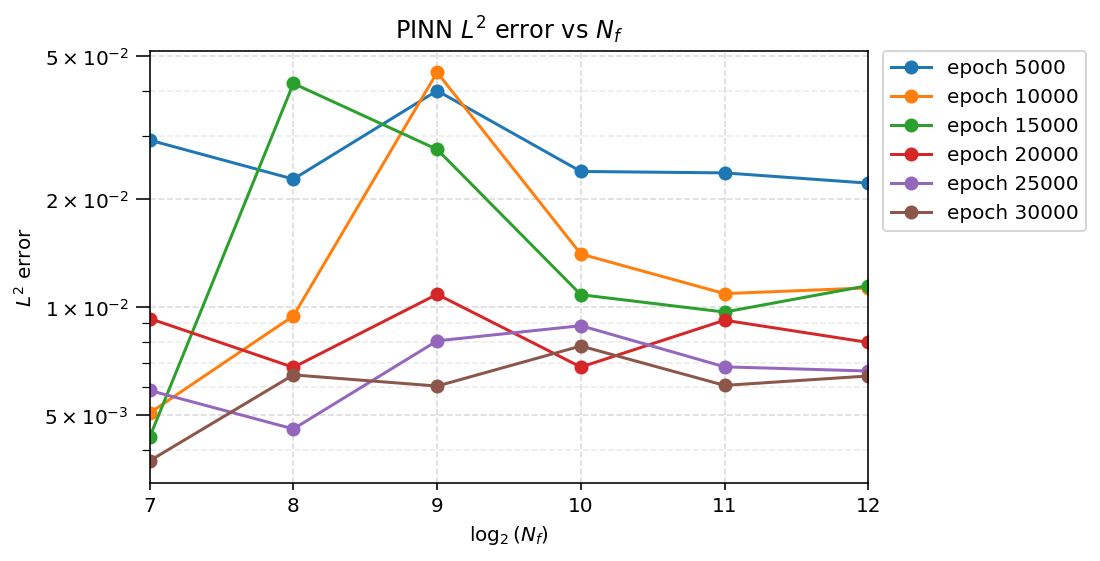}
    \caption{ $L^\infty$ error (left) and $L^2$ error (right) of the PINN with different epoch number with tanh activation vs. number of interior points $N_f$ (on log scale).}
    \label{fig:pinn_interior_points}
\end{figure}

{Figure \ref{fig:pinn_interior_points} shows the convergence of the PINN solution when the number of interior points $N_f$ increases for an MLP with $4$ hidden layers with $64$ neurons in each case. We observed that $N_f=2^9$ seems to be sufficient (large enough such that further refinement does not improve the accuracy significantly) in this setting, and therefore, we use this number in later computations.}

\begin{figure}[h!]
    \centering
    \includegraphics[width=0.5\textwidth,height=.35\textwidth]{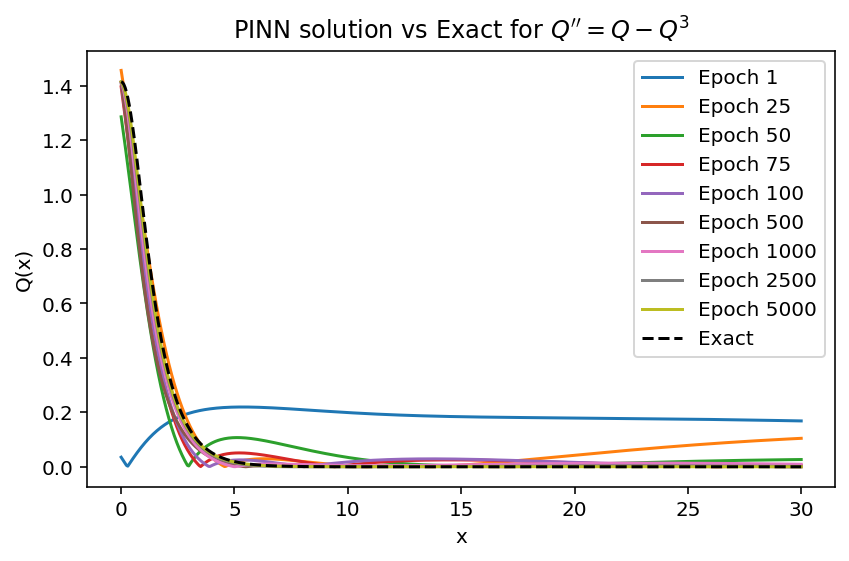}
    \includegraphics[width=0.48\textwidth,height=.35\textwidth]{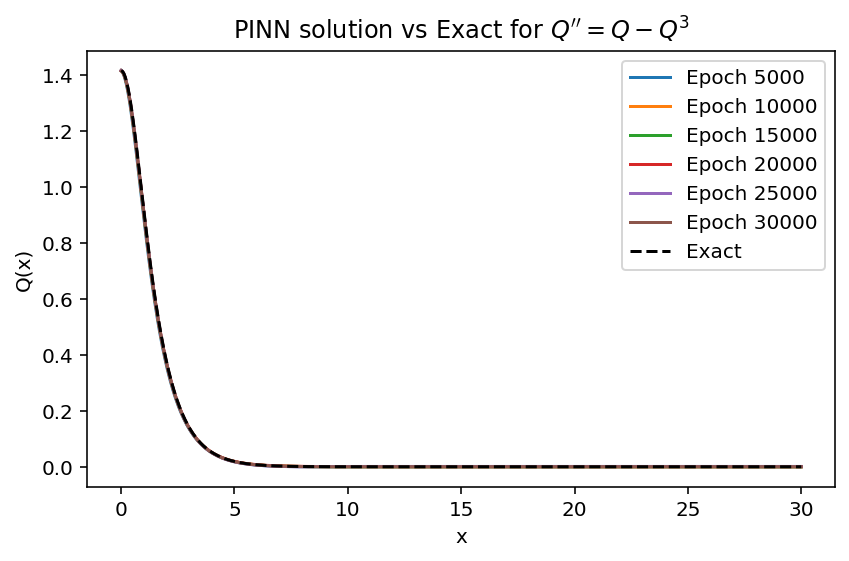}
    \caption{PINN profiles varying with number of epochs (via $\tanh$ activation).}
    \label{fig:PINN_profile}
\end{figure}

In Figure \ref{fig:PINN_profile} we show a comparison of profiles depending on the number of epochs: on the left plot we start with 1 epoch and go up to 5,000, comparing them with the exact solution; on the right plot we go from 5,000 epochs up to 30,000 epochs and also compare with the exact solution. While one can see the difference in computing the profiles visually in the left plot, it is almost not visible on the right plot, however, from Figures \ref{fig:erroranalysis_PINN}-\ref{fig:pinn_interior_points} we infer that the error (either in $L^\infty$ or $L^2$ norms) is on the order of $10^{-2}$.

\begin{figure}[h!]
    \centering
    \includegraphics[width=0.49\textwidth,height=0.35\textwidth]{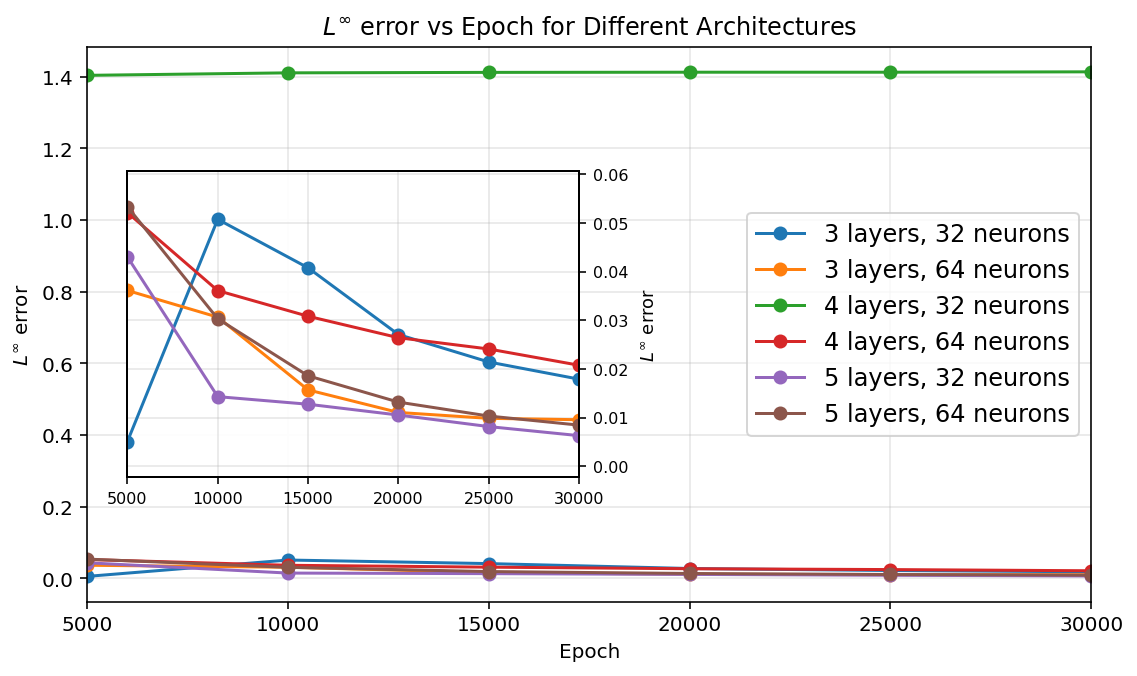}
       \includegraphics[width=0.49\textwidth,height=0.35\textwidth]{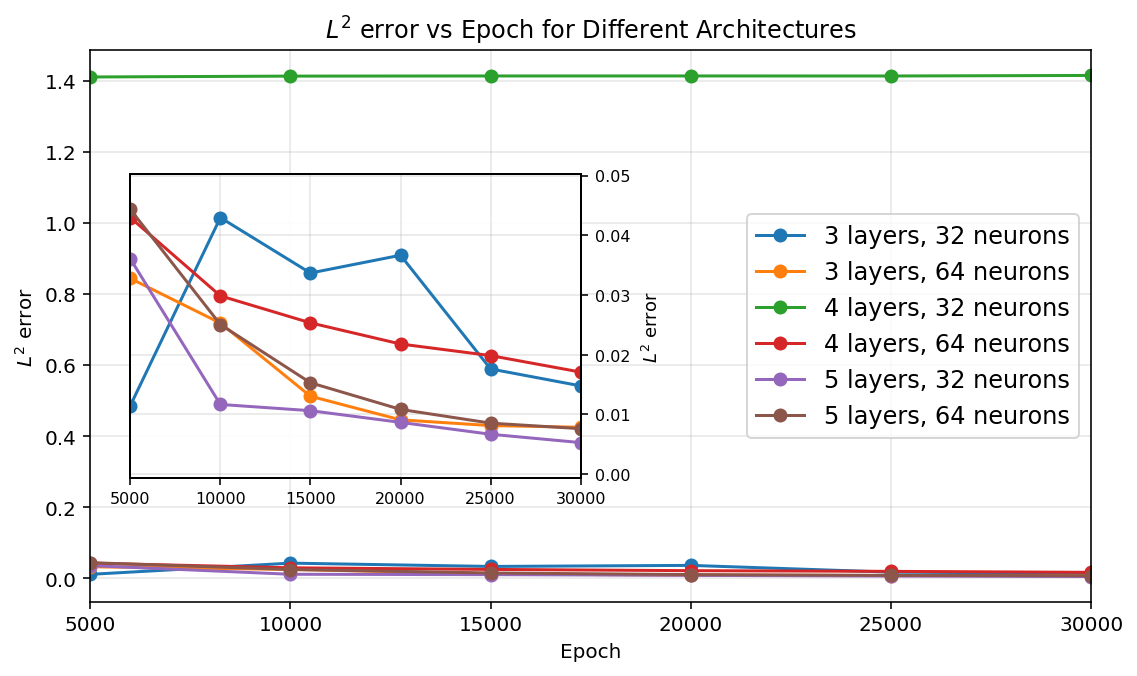}
    \caption{$L^\infty$ error (left) and $L^2$ error (right) of the PINN solution for \eqref{2ndODE-finite} with SiLU activation function vs. the number of epochs.}
    \label{fig:erroranalysis_PINN_silu}
\end{figure}

\begin{figure}[h!]
    \centering
    \includegraphics[width=0.49\textwidth,height=0.33\textwidth]{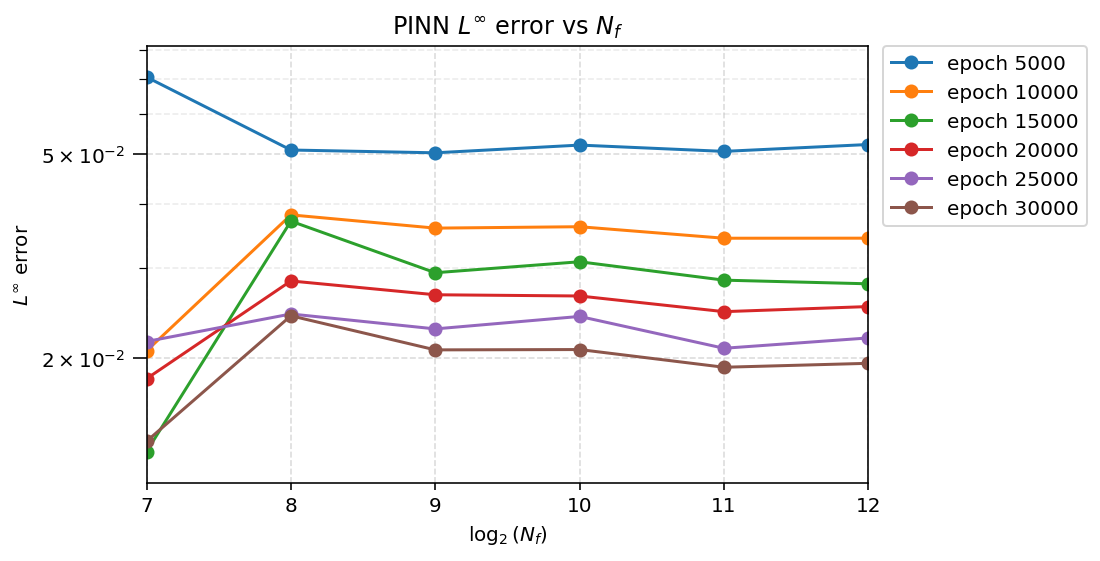}
       \includegraphics[width=0.49\textwidth,height=0.33\textwidth]{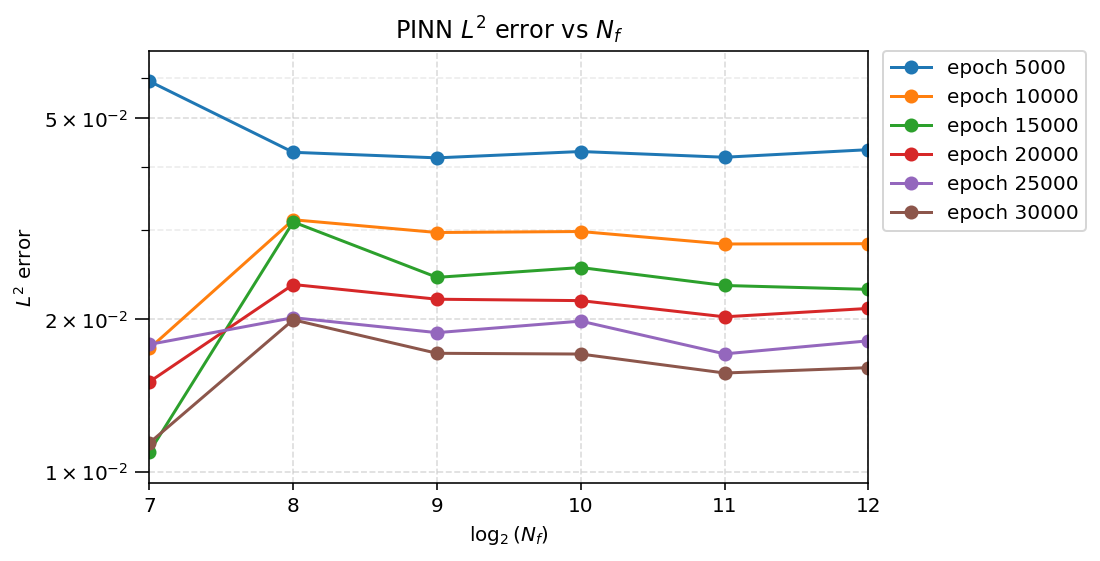}
           \caption{ $L^\infty$ error (left) and $L^2$ error (right) of the PINN solution with SiLU activation function vs. the number of interior points $N_f$ (on a log scale).}
    \label{fig:pinn_interior_points_silu}
\end{figure}

\begin{figure}[h!]
    \centering
    \includegraphics[width=0.49\textwidth, height=.33\textwidth]
    {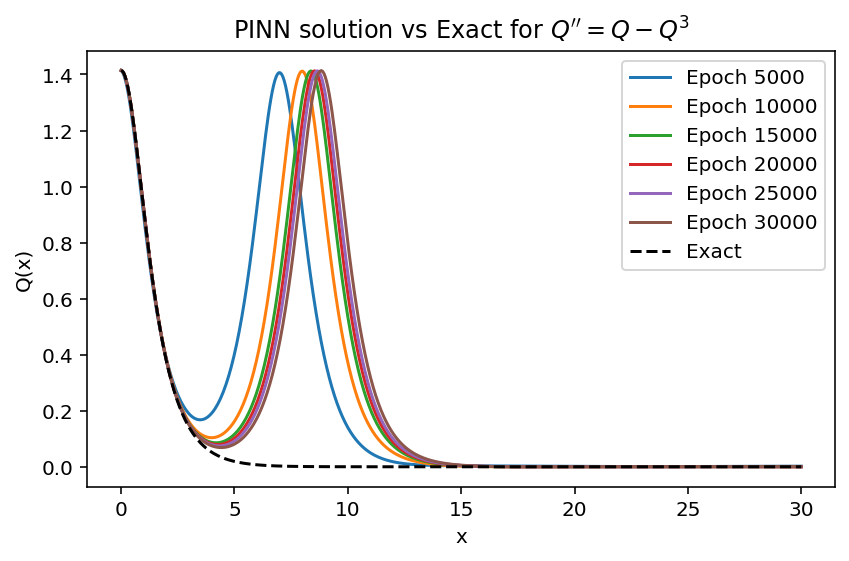}
    \includegraphics[width=0.49\textwidth, height=.33\textwidth]{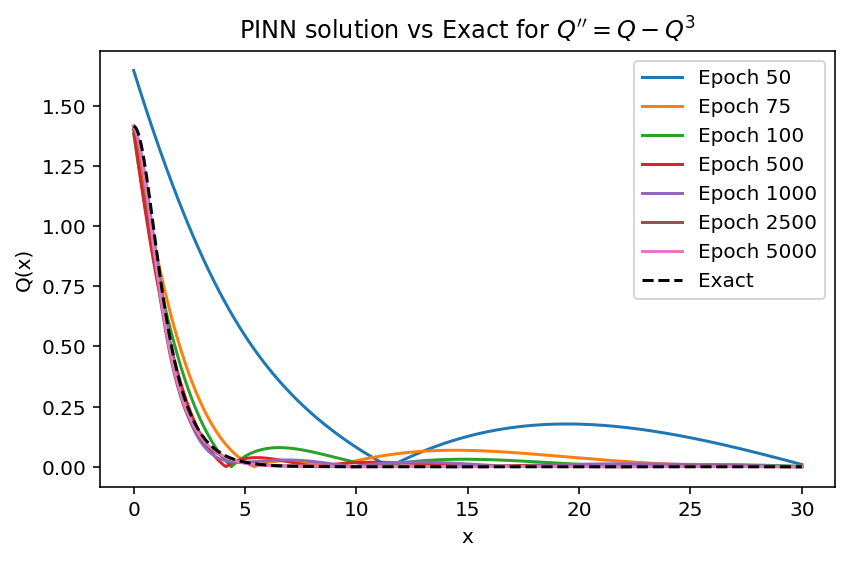}
    \caption{PINN solution profiles using the SiLU activation at increasing training epochs for a 4-layer network with 32 neurons (left) and 64 neurons (right), compared against the exact profile.}
\label{fig:silu4l32n}
\end{figure}

\newpage
We next consider a different activation function, $\mathrm{SiLU}(x) = x/(1+e^{-x})$. Using the same settings from Table \ref{tab:pinn_hyperparams} as we did for the $\tanh$ activation, we report findings in Figures \ref{fig:erroranalysis_PINN_silu} ($L^\infty$ and $L^2$ errors vs. epoch number), \ref{fig:pinn_interior_points_silu} ($L^\infty$ and $L^2$ errors vs. number of points $N_f$), \ref{fig:silu4l32n} (convergence of profiles).


One can notice that one of the architectures (4 layers, 32 neurons) in Figure \ref{fig:erroranalysis_PINN_silu} has a significantly large error, 
we investigate the profile convergence in the left plot of Figure \ref{fig:silu4l32n}.
In this specific example (4 layer, 32 neurons), the PINN with the SiLU activation function is not able to detect correctly and completely the necessary profile, see left plot of Figure \ref{fig:silu4l32n}, with only the initial decay and the decay in the tail matching the exact solution; for $3<x<15$ one can observe that it generates another hump, which does not seem to disappear with epochs increased, and does not correspond to the positive monotone (in $r = |x|$ coordinate) ground state profile 
(though, this might be useful in computation of branching profiles as in \cite{YRZ2019} or excited and bound state profiles).

Note that if in the same 4-layer network the number of neurons is doubled from 32 to 64, the PINN solution has a better convergence to the exact solution with the epoch number increasing as shown in the right plot of Figure \ref{fig:silu4l32n}. One can compare this with a similar convergence in the left plot of Figure \ref{fig:PINN_profile} with the $\tanh$ activation function, where the number of epochs is changing in a similar range. Therefore, we conclude that the SiLU activation may not always be a suitable candidate for the activation function for an ODE solution computation via PINNs, or several architectures would have to be used to find the appropriate solution.  

In Table \ref{tab:pinn_p_concise} we provide specific values for the loss, CPU time, the $L^\infty$ and $L^2$ errors to compare these two activation functions. One notices that SiLU gives slightly inferior numbers, in particular, the CPU time is longer and the errors are a bit higher. We next discuss the computation of solitary wave profiles for different powers $p$.  
\begin{table}[H]
\centering
\scriptsize
\setlength{\tabcolsep}{3pt}
\renewcommand{\arraystretch}{1.05}

\begin{tabular}{|c||c|c|c|c||c|c|c|c|}
\hline
 & \multicolumn{4}{c||}{PINN with $\tanh$ activation} & \multicolumn{4}{c|}{PINN with SiLU activation} \\
\hline\
$p$ & Loss & CPU time & $L^\infty$ error & $L^2$ error & Loss & CPU time & $L^\infty$ error & $L^2$ error \\
\hline\hline
$10/9$ & 2.801825e-07 & 1817.3906 & 2.905399e-02 & 2.073294e-02 & 5.874724e-08 & 2495.7969 & 1.405067e-02 & 1.008419e-02 \\
\hline
$16/9$ & 1.162162e-07 & 1718.0938 & 1.407494e-02 & 1.070179e-02 & 1.612241e-07 & 2331.1094 & 2.169832e-02 & 1.651699e-02 \\
\hline
$2$    & 1.401239e-05 & 1644.0781 & 9.877680e-03 & 8.691553e-03 & 3.810378e-08 & 2273.3438 & 2.099043e-02 & 1.627930e-02 \\
\hline
$3$    & 1.216845e-05 & 1576.0000 & 8.875633e-03 & 7.784941e-03 & 3.858329e-07 & 2388.8594 & 2.076610e-02 & 1.708686e-02 \\
\hline
$4$    & 4.361955e-08 & 1508.5625 & 6.420298e-03 & 5.453788e-03 & 3.072228e-07 & 2318.4688 & 1.535378e-02 & 1.312327e-02 \\
\hline
$4.5$  & 1.761227e-07 & 1690.2812 & 6.465403e-03 & 5.551418e-03 & 4.239125e-07 & 2343.9375 & 1.436826e-02 & 1.242997e-02 \\
\hline
$5$    & 8.773534e-08 & 1572.6562 & 5.856099e-03 & 5.055659e-03 & 3.601354e-07 & 2363.6719 & 1.315570e-02 & 1.158138e-02 \\
\hline
$7$    & 6.893825e-08 & 1642.9531 & 6.049440e-03 & 5.461686e-03 & 8.585262e-07 & 2376.9844 & 7.738219e-03 & 7.052342e-03 \\
\hline
\end{tabular}
\caption{Comparison of PINN performance with $\tanh$ and SiLU activations, 30,000 epochs, $N_{f}=2^{10}$, $L=30$, 4 hidden layers and 64 neurons for $Q''=Q-Q^p$.}
\label{tab:pinn_p_concise}
\end{table}
\subsubsection{General power $p$}
After verifying the PINN solution for the cubic case  $p=3$ and $b=1$, $\gamma=1$, we now test a general nonlinearity power $p$
\begin{equation*}
    Q''=bQ-\gamma Q^p.
\end{equation*}
Table \ref{tab:pinn_p_concise} shows the convergence of PINN solution for a variety of powers
$$
p=\tfrac{10}9,\, \tfrac{16}9,\,2,\,3,\,4,\,4.5,\,5,\,7,
$$
where we included powers close to 1, subcritical powers, critical and supercritical ones. 

With an MLP of $4$ hidden layers and $64$ neurons each, we first observe that the order 
of the $L^\infty$ and $L^2$ errors have comparable accuracy with the finite difference method as shown in Tables \ref{tab:NP_p_tenninth}-\ref{tab:NP_k1p1p3} and Tables \ref{tab:appendix_1}-\ref{tab:appendix_end} in Appendix (PINNs with both activation functions performing less accurate for smaller $p$, and SiLU is performing slightly worse than the tanh activation overall). In terms of these norm errors, the best performing is the classical Petviashvili method.

Secondly, and perhaps more importantly, is that the CPU time is drastically different when PINNS compared to the classical methods: at least on the order of $10^{-1}$ (for FD+Newton) and $10^{-2} \div 10^{-8}$ (for Petviashvili) vs. $2 \times 10^{3}$ (for PINNs), thus, at least $10^{4} \div 10^{11}$ times slower (in this one dimensional setting). Nevertheless, the PINNs approach maybe more preferable in other settings, see next subsection; 
we also show below how operator learning can be done fast, once training on the data is done (which can, in its turn, take some time).   

\subsubsection{Initial conclusions about PINNs vs. classical schemes computations of solitary wave profiles}
\label{rem:PINN}
We now can make the following remarks:

1. {\bf (CPU time)} When comparing the CPU time, the PINN algorithm takes considerably more CPU time than the classical methods (in this 1D setting). 

However, one can use a smaller MLP and less number of epochs if highly accurate solutions are not needed, which will reduce the CPU time.

Secondly, one notices that the classical numerical methods require a `good' enough initial guess (otherwise, for example, a constant zero is a solution and can appear as one of the possible results from the classical methods). PINN is an optimization algorithm and does not require a `good' initial guess, which increases the computational time in finding the appropriate solution. 

The CPU time can change with better computational capabilities, see next section about that. 
\smallskip

2. {\bf (No discretization or mesh generation)} One advantage of the PINN method is that it learns to satisfy both the ODE and the boundary conditions simultaneously, without requiring explicit discretization or mesh generation.
\smallskip

3. {\bf (Transfer learning/Operator learning)} If 
a family of equations of the form \eqref{2ndODE} needs to be solved for different combinations of parameters $(b,\gamma,p)$, then for moderately sized domains (e.g., $L=30$), {\it transfer learning} can be employed to train the PINN on a representative parameter set, and then subsequently, reuse or fine-tune the pre-trained model for nearby parameter values; see, for example, \cite{WBRAZRL25}, and we discuss a more general {\it operator learning} in Section \ref{S:OL}.

\subsubsection{Comparison with the ``secondary" computational device}\label{S:secondary}

The runs on ``Secondary" device were performed on a 64-bit machine with an AMD Ryzen 5 7535HS (3.30 GHz, 6 cores/12 threads), integrated Radeon graphics (no CUDA-capable GPU), and 16 GB RAM (15.3 GB usable).
To compare, we recall that all previous runs were performed on the ``Primary" device, which is a 64-bit machine with an AMD Ryzen 5 7530U (2.00 GHz) processor, 12 GB RAM (11.4 GB usable), x64-based architecture. 

As we show in Table \ref{tab:computercomparison}, the secondary system (Ryzen 5 7535HS, 3.30 GHz, 16 GB RAM) will generally give better performance in CPU time than the primary system (Ryzen 5 7530U, 2.00 GHz, 12 GB RAM), especially for CPU-intensive Python/PyTorch or NumPy code and larger models or datasets, thanks to its higher clock speed and extra memory; both are 64-bit x64 machines using CPU-only computation (no CUDA-capable GPU), therefore, the main difference in runtime comes from the stronger CPU and larger RAM on the secondary system. Thus, comparing the columns in Table \ref{tab:computercomparison} one can observe that an increase in CPU frequency by 65\% in GHz and 35\% in RAM capacity gives about 25-30\% faster CPU time, which maintaining the same errors and loss. 

\begin{table}[htb!]
\centering
\setlength{\tabcolsep}{3pt}
\renewcommand{\arraystretch}{1.05}
\begin{tabular}{|c||c|c|c|c||c|c|c|c|}
\hline
 & \multicolumn{4}{c||}{Primary Device} & \multicolumn{4}{c|}{Secondary Device} \\
\hline\hline
Epoch & Loss & CPU time & $L^\infty$ error & $L^2$ error & Loss & CPU time & $L^\infty$ error & $L^2$ error \\
\hline\hline
5{,}000  & 3.342e-06 & 319.27  & 2.898e-02 & 2.385e-02 & 3.342e-06 & 223.56  & 2.898e-02 & 2.385e-02 \\
\hline
10{,}000 & 7.666e-07 & 630.95  & 1.717e-02 & 1.404e-02 & 7.666e-07 & 450.97  & 1.717e-02 & 1.404e-02 \\
\hline
15{,}000 & 1.932e-07 & 914.50  & 1.320e-02 & 1.082e-02 & 1.932e-07 & 676.58  & 1.320e-02 & 1.082e-02 \\
\hline
20{,}000 & 9.857e-07 & 1201.94 & 8.312e-03 & 6.802e-03 & 9.857e-07 & 901.23  & 8.312e-03 & 6.802e-03 \\
\hline
25{,}000 & 8.123e-06 & 1490.64 & 9.003e-03 & 8.869e-03 & 8.123e-06 & 1131.53 & 9.003e-03 & 8.869e-03 \\
\hline
30{,}000 & 1.217e-05 & 1783.22 & 8.876e-03 & 7.785e-03 & 1.217e-05 & 1367.78 & 8.876e-03 & 7.785e-03 \\
\hline
\end{tabular}
\caption{PINN device comparison with $L=30$, $N=2^{10}$, $\tanh$ activation, 4 hidden layers and 64 neurons for $Q''=Q-Q^3$.}
\label{tab:computercomparison}
\end{table}

\subsection{Operator Learning:   Deep Operator Network and Fourier Neural Operator}\label{S:OL}
While PINNs provide a flexible, mesh-free framework, their performance in stationary wave problems is often limited by the need to repeatedly solve a challenging optimization problem for each new set of parameters, with accuracy and efficiency sensitive to the choice of collocation points and loss weighting. In contrast, operator learning methods aim to approximate the solution operator directly, allowing a single trained model to generalize across a family of input parameters or forcing terms. Approaches such as  Operator Networks (DeepONet) \cite{lu2021deeponet} and Fourier Neural Operators (FNO) \cite{FNO,li2021fourier} offer a potentially more efficient alternative for parametric studies (which can be very valuable in control theory and such) by enabling rapid inference for new problem instances at the cost of single model training.

The goal of operator learning for \eqref{2ndODE-finite} and \eqref{2ndODEBC-Dirichlet} is to  learn the mapping
\[
\mathcal{G}:\ (b,\gamma,p)\ \longmapsto\ Q(\cdot;b,\gamma,p)
\]
\begin{equation}
Q''(x)=b\,Q(x)-\gamma\,Q(x)^p,\qquad
Q(0)=Q_{exact}(0,b,\gamma,p),\quad Q(L)=0,\quad x\in[0,L]. \label{Operator}
\end{equation}
Here, \(b\in[b_{min},b_{max}]\), \(\gamma\in[\gamma_{min},\gamma_{max}]\), \(p\in[p_{min},p_{max}]\),  and \(L=30\) is large enough that \(Q(L)\) is effectively zero. We test two such operator learnings. 

\subsubsection{Deep Operator Network (DeepONet)} To learn the  operator setting \eqref{Operator},  DeepONet uses two  MLP networks
\begin{align*}
\text{Branch: } & B_\theta:(b,\gamma,p)\mapsto\mathbb{R}^m,\\
\text{Trunk: }  & T_\theta:x\mapsto\mathbb{R}^m,
\end{align*}
to  predict solution by an inner product
\[
Q_\theta(x;b,\gamma,p) \;=\; B_\theta(b,\gamma,p)^\top T_\theta(x)
\;=\; \sum_{k=1}^m B_{\theta,k}(b,\gamma,p)\,T_{\theta,k}(x).
\]
The details of the implementation are shown in Algorithm \ref{alg:deeponet-parametric}.  
After training, for any new triple $(b,\gamma,p)$ in the parameter range,
the approximate solution is given by
\[
Q_\theta(x;b,\gamma,p) = B_\theta(b,\gamma,p)^\top T_\theta(x),\qquad x\in[0,L],
\]
evaluated on any desired spatial grid.
\smallskip

\begin{algorithm}[h]
\caption{Parametric DeepONet solving $Q'' = bQ - \gamma Q^p$ on $[0,L]$}
\label{alg:deeponet-parametric}
\begin{algorithmic}[1]
\Require Parameter ranges
$(b,\gamma,p)\in[b_{\min},b_{\max}]\times[\gamma_{\min},\gamma_{\max}]\times[p_{\min},p_{\max}]$;
training grid $\{x_j\}_{j=1}^M\subset[0,L]$;
collocation set $\{x_k\}_{k=1}^{M_c}\subset[0,L]$;
total epochs $E$; domain length $L$; learning rate $\eta$; weight $\lambda_{\mathrm{bc}}$, etc.

\State Initialize branch MLP $B_\theta:\mathbb{R}^3\to\mathbb{R}^m$ and trunk MLP $T_\theta:[0,L]\to\mathbb{R}^m$
with parameters $\theta$, optimizer, etc.

\For{epoch $=1$ to $E$}
  \State Sample a mini-batch of parameter triples
    $\{(b_i,\gamma_i,p_i)\}_{i=1}^B$ from the parameter domain.
  \For{each $(b_i,\gamma_i,p_i)$ in the batch}
    \State Compute branch features
      $c_i = B_\theta(b_i,\gamma_i,p_i)\in\mathbb{R}^m$.
    \State For all $x_j$, compute trunk features
      $t_j = T_\theta(x_j)\in\mathbb{R}^m$.
    \State Form predictions
      $Q_\theta(x_j;b_i,\gamma_i,p_i) = c_i^\top t_j$ for $j=1,\dots,M$.
    \State Use automatic differentiation w.r.t.\ $x$ to obtain
      $Q'_\theta(x;b_i,\gamma_i,p_i)$ and $Q''_\theta(x;b_i,\gamma_i,p_i)$
      at $x\in\{x_k\}$.
  \EndFor
  \State Define the physics loss  as
    \[
    \mathcal{L}_{\mathrm{phys}}
    = \frac{1}{B\,M_c}\sum_{i=1}^B\sum_{k=1}^{M_c}
      \Big(
        Q''_\theta(x_k;b_i,\gamma_i,p_i)
        - \big[b_i\,Q_\theta(x_k;b_i,\gamma_i,p_i)
                - \gamma_i\,Q_\theta(x_k;b_i,\gamma_i,p_i)^{p_i}
          \big]
      \Big)^2.
    \]
  \State Evaluate left exact boundary targets
    $Q_{\mathrm{L},i}^{\mathrm{exact}} = Q_{\mathrm{exact}}(0;b_i,\gamma_i,p_i)$.
  \State Define boundary loss
    \[
    \mathcal{L}_{\mathrm{bc}}
    = \frac{1}{B}\sum_{i=1}^B
      \Big(
        Q_\theta(0;b_i,\gamma_i,p_i) - Q_{\mathrm{L},i}^{\mathrm{exact}}
      \Big)^2
      +
      \frac{1}{B}\sum_{i=1}^B
      \Big(
        Q_\theta(L;b_i,\gamma_i,p_i)
      \Big)^2.
    \]
  \State Total loss:
    $\mathcal{L} = \mathcal{L}_{\mathrm{phys}} + \lambda_{\mathrm{bc}}\,\mathcal{L}_{\mathrm{bc}}$.
  \State Update $\theta \leftarrow \theta - \eta \nabla_\theta \mathcal{L}$
\EndFor
\State \textbf{Output:} trained DeepONet \(Q_\theta(x;b,\gamma,p)\).
\end{algorithmic}
\end{algorithm}
\begin{table}[h!]
\centering
\setlength{\tabcolsep}{3pt}
\renewcommand{\arraystretch}{1.05}
\begin{tabular}{|c||c|c||c|c|}
\hline
 & \multicolumn{2}{c||}{$p_{\text{train}}\in[2,6]$} & \multicolumn{2}{c|}{$p_{\text{train}}\in[10/9,7]$} \\
\hline\hline
$p$ & $L^\infty$ error & $L^2$ error & $L^\infty$ error & $L^2$ error \\
\hline\hline
$10/9$ & 1.029928e+00 & 2.244082e+00 & 6.175135e-01 & 1.306146e+00 \\
\hline
$16/9$ & 1.839474e-01 & 2.525783e-01 & 3.920233e-02 & 5.015318e-02 \\
\hline
$2$    & 1.110107e-01 & 1.352280e-01 & 4.297459e-02 & 4.786296e-02 \\
\hline
$3$    & 4.082304e-02 & 5.077833e-02 & 2.482951e-02 & 4.137911e-02 \\
\hline
$4$    & 5.400246e-02 & 6.349091e-02 & 5.896181e-02 & 6.682276e-02 \\
\hline
$4.5$  & 3.906476e-02 & 4.181148e-02 & 6.312329e-02 & 6.690990e-02 \\
\hline
$5$    & 5.147314e-02 & 4.804274e-02 & 4.854238e-02 & 3.214040e-02 \\
\hline
$7$    & 8.846533e-02 & 7.545136e-02 & 8.498812e-02 & 6.473307e-02 \\
\hline
\end{tabular}
\caption{DeepONet performance across $p$ for two $p$ training ranges, with $M_c=2^{10}$, $L=30$, $N_{train}=40$, $N_{test}=10$, 2 hidden layers, 64 neurons, modes, and $E=5{,}000$ for $Q''=Q-Q^p$. Training times: $CPU_{\text{train}}=62{,}702.7031$ sec for $p_{\text{train}}\in[2,6]$ and $CPU_{\text{train}}=42{,}653.1094$ sec for $p_{\text{train}}\in[10/9,7]$.}
\label{tab:DeeONet}
\end{table}
Since the power $p$ plays the major role in the solitary wave profile, we show in Table \ref{tab:DeeONet} the $L^\infty$ and $L^2$ inference errors for $p=10/9,\,16/9,\,2,\,3,\,4,\,4.5,\,5,\,7$ after training on $p\in[2,6]$ (left) and $p\in[10/9,7]$ with fixed $b=\gamma=1$.  In general, we obtain $O(10^{-2})$ inference errors, even when the conferencing $p$ is not in the training range (left). The errors are comparable to the PINN errors in Table \ref{tab:computercomparison} for $p=3$. It is also observed that the inference error increases as $p\rightarrow 1^+$, 
which is expected, since the solution operator becomes nearly singular and dominated by a global amplitude mode. This leads to severe ill-conditioning and a loss of separability in DeepONet’s branch–trunk representation, thereby significantly degrading training stability and generalization performance.
\smallskip

In general DeepONet provides mesh-free, almost-instant inference and cross-instance generalization, at the expense of a high upfront training cost but inexpensive evaluation.

\begin{Remark}

\begin{itemize}
\item[1.] The collocation set $\{x_k\}_{k=1}^{M_c}$ is  chosen as a subset (randomly)
of the training grid $\{x_j\}_{j=1}^M$ in the simulation. 
\item[2.] Automatic differentiation ($torch.autograd$ in PyTorch) is used in DeepONet to evaluate $Q_\theta'$ and $Q_\theta''$ at collocation points. 

\item[3.]  Additional supervised data loss term  can also be added to the loss function $\mathcal{L}$ for the DeepONet if solutions are known for some specific parameters $(b_s,\gamma_s,p_s)$ or at some special locations.
\end{itemize}
   
\end{Remark}

\subsubsection{Fourier Neural Operator (FNO)}
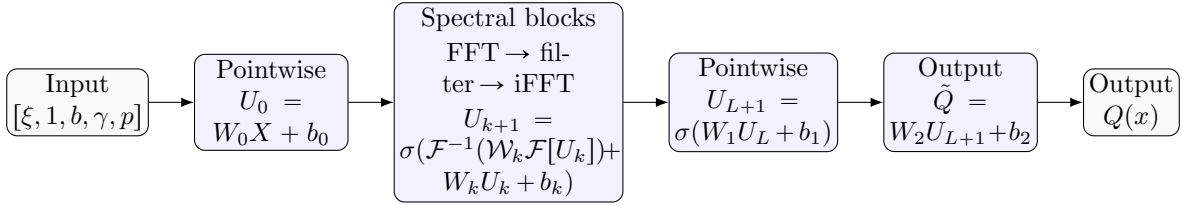
\begin{figure}[h!] 
\centering
\begin{tikzpicture}[
  font=\small,
  >=Latex,
  node distance=6mm and 6mm,
  box/.style={draw, rounded corners, minimum height=8mm, inner sep=2pt, fill=blue!5, align=center},
  node/.style={draw, rounded corners, minimum height=8mm, inner sep=2pt, fill=gray!5, align=center}
]

\node[node] (inp)
{Input \\ $[\xi,1,b,\gamma,p]$};

\node[box, right=of inp, text width=1.9cm] (proj0)
{Pointwise \\ $U_0=W_0X+b_0$};

\node[box, right=of proj0, text width=2.9cm] (spec)
{Spectral blocks \\[1pt]
FFT $\!\to$ filter $\!\to$ iFFT \\[1pt]
$U_{k+1}\!=\!\sigma(\mathcal{F}^{-1}(\mathcal{W}_k\mathcal{F}[U_k])\!+\!W_kU_k\!+\!b_k)$};

\node[box, right=of spec, text width=2.1cm] (proj1)
{Pointwise \\ $U_{L+1}=\sigma(W_1U_L+b_1)$};

\node[box, right=of proj1, text width=1.9cm] (proj2)
{Output \\ $\tilde Q=W_2U_{L+1}+b_2$};

\node[node, right=of proj2]
(out) {Output \\ $ Q(x)$};

\draw[->] (inp) -- (proj0);
\draw[->] (proj0) -- (spec);
\draw[->] (spec) -- (proj1);
\draw[->] (proj1) -- (proj2);
\draw[->] (proj2) -- (out);

\end{tikzpicture}
\caption{Network architecture of Fourier Neural Operator $\mathrm{FNO}(X;\theta)$}
\label{fig:FNO_arch}
\end{figure}

The goal of the parametric  (FNO) \cite{FNO,li2021fourier} is to  learn the same operator \eqref{Operator}.  We set the input of FNO as
$[\,\xi,\,1,\,b,\,\gamma,\,p\,]$, 
where the spatial $\xi$ are partitioned uniformly in $[0,1]$ and constant fields  \((b,\gamma,p)\) are drawn uniformly from the parameter domains. We show  the details of implementation in Algorithm \ref{alg:parametric-fno}, with the FNO architecture in the ``Forward pass" (line 9, Algorithm \ref{alg:parametric-fno}) shown in Figure \ref{fig:FNO_arch} for reader's convenience. 
After training, we obtain an FNO network ${Q}(x;b,\gamma,p)$, which is used to inference a predicted solution $Q_{\text{pred}}(x)$ for specified parameters \((b,\gamma,p)\) within (or near) the training parameter domain. 

\begin{algorithm}[h]
\caption{Parametric FNO solving $Q'' = bQ - \gamma Q^p$ on $[0,L]$}
\label{alg:parametric-fno}
\begin{algorithmic}[1]
\Require
  Parameter ranges $b\in[b_{\min},b_{\max}]$, $\gamma\in[\gamma_{\min},\gamma_{\max}]$, $p\in[p_{\min},p_{\max}]$;
  domain length $L$; grid size $N$;
  batch size $B$; total epochs $E$;
  learning rate $\eta$;
  weights $w_{\mathrm{bc}}$,etc.

\State \textbf{Initialize:}
  Construct FNO: width $W$, $L_{\mathrm{FNO}}$ spectral layers, parameters $\theta$, optimizer, etc.
\For{$epoch = 1,\cdots, E$}
  \State Sample $B$ random parameter triples
    $(b_i,\gamma_i,p_i)$ from the training domain.
  \For{each $i=1,\cdots,B$}
    \State Discretize spatial grid $x_j=j\Delta x$, $\Delta x=L/(N-1)$.
    \State Build input tensor
      $X_i(x_j) = [\,x_j/L,\,1,\,b_i,\,\gamma_i,\,p_i\,]$.
    \State Compute exact left boundary $Q_{exact}(0,b_i,\gamma_i,p_i)$.
  \EndFor
  \State Forward pass: $Q_i(x_j) = \mathrm{FNO}(X_i(x_j);\theta)$ for all $i$.
  \State Compute discrete Laplacian
      $(\mathcal{L}_h Q_i)_j = (Q_{i,j-1}-2Q_{i,j}+Q_{i,j+1})/\Delta x^2$.
  \State Evaluate interior residual
      $\mathcal{R}_{i,j} = (\mathcal{L}_h Q_i)_j - (b_i Q_{i,j} - \gamma_i Q_{i,j}^{p_i})$.
  \State Form losses:
    \begin{align*}
      \mathcal{L}_{\mathrm{phys}} \gets \tfrac{1}{B(N-1)}\sum_{i,j}\mathcal{R}_{i,j}^2, \quad
      \mathcal{L}_{\mathrm{bc}} &\gets \tfrac{1}{B}\sum_i[(Q_{i,0}-Q_{exact}(0,b_i,\gamma_i,p_i))^2+(Q_{i,N-1})^2].
    \end{align*}
  \State Compute total loss
    $\mathcal{L}=\mathcal{L}_{\mathrm{phys}}+w_{\mathrm{bc}}\mathcal{L}_{\mathrm{bc}}$.
  \State Back-propagate and update parameters $\theta\leftarrow\theta-\eta\nabla_\theta \mathcal{L}$.
\EndFor
\State \textbf{Output:} Trained FNO  \(Q_\theta(x;b,\gamma,p)\).
\end{algorithmic}
\end{algorithm}

\begin{figure}[htp!]
    \centering
    \includegraphics[width=0.49\textwidth,height=.37\textwidth]{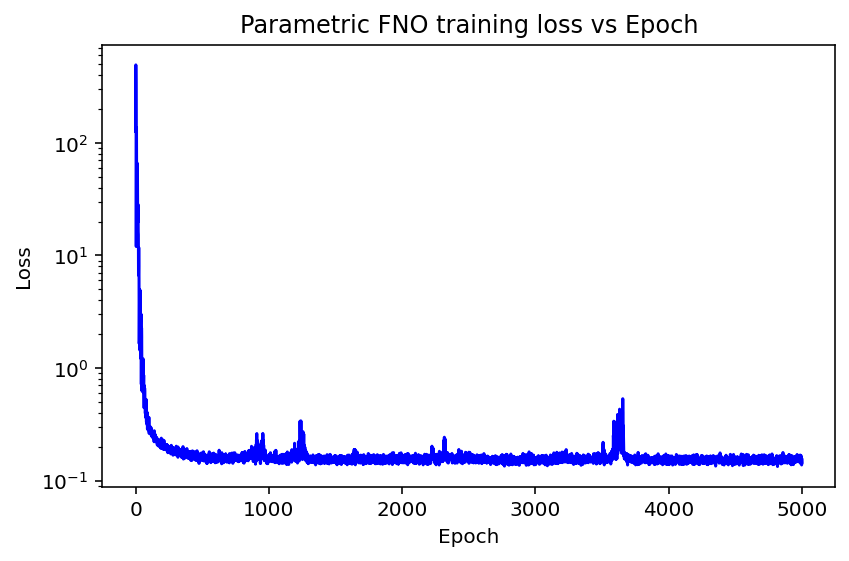}
    \includegraphics[width=0.49\textwidth,height=.37\textwidth]{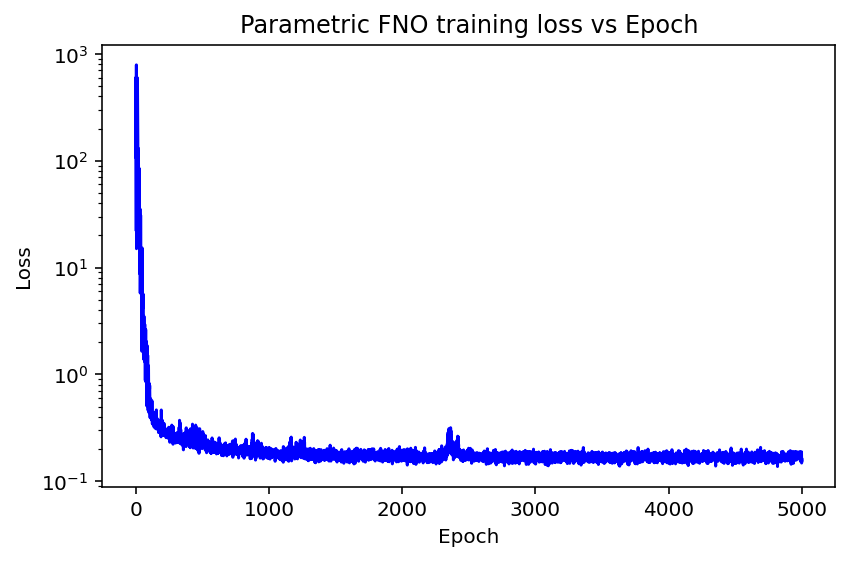}
    \caption{Loss vs Epoch for FNO for $p_{train}\in[2, 6]$ (left) and $p_{train}\in[10/9, 7]$ (right), $b_{train} =1$, $N=2^{10}$, $L=30$, $W=64$, $L_{FNO}=4$, $B=8$, $lr=3 \times 10^{-3}$, and $E=5,000$ for $Q''=bQ-Q^p$. }
    \label{fig:FNO_epoch}
\end{figure}

\begin{table}[h!]
\centering
\setlength{\tabcolsep}{3pt}
\renewcommand{\arraystretch}{1.05}
\begin{tabular}{|c||c|c||c|c|}
\hline
 & \multicolumn{2}{c||}{$p_{\text{train}}\in[2,6]$} & \multicolumn{2}{c|}{$p_{\text{train}}\in[10/9,7]$} \\
\hline\hline
$p$ & $L^\infty$ error & $L^2$ error & $L^\infty$ error & $L^2$ error \\
\hline\hline
$10/9$ & 1.053121e+00 & 8.435218e-01 &1.107145e+00 & 8.851508e-01 \\ \hline
$16/9$ & 5.687321e-01 & 3.587534e-01 & 5.876326e-01 & 4.187138e-01 \\ \hline
$2$    & 3.377558e-01 & 2.392224e-01 & 4.741016e-01 & 3.059828e-01 \\\hline
$3$    & 3.486853e-01 & 1.905931e-01  & 3.232563e-01 & 1.888838e-01 \\\hline
$4$    & 4.634934e-01 & 1.872775e-01 &  4.777224e-01 & 1.804326e-01 \\\hline
$4.5$  & 5.165036e-01 & 1.999541e-01 & 5.074022e-01 & 1.919678e-01\\\hline
$5$    & 5.693246e-01 & 2.048454e-01 & 5.651524e-01 & 2.097748e-01 \\\hline
$7$    & 7.610273e-01 & 2.371290e-01 & 7.462914e-01 & 2.453445e-01 \\\hline
9  & 6.886868e-01 & 2.270165e-01 & 5.504158e-01 & 3.604365e-01  \\ \hline
11  & 5.528975e-01 & 1.568201e-01  & 8.165537e-01 & 6.160263e-01 \\ \hline
25 & 2.125769e+00 & 1.283375e+00 & 3.186769e+00 & 2.517480e+00\\ \hline
\end{tabular}
\caption{FNO performance across $p$ for two $p$ training ranges, with $N=2^{10}$, $L=30$, $W=64$, $L_{FNO}=4$, $B=8$, $lr=5\times10^{-4}$, and $E=5{,}000$ for $Q''=Q-Q^p$. Training times: $CPU_{\text{train}}=1{,}925.1875$ sec for $p_{\text{train}}\in[2,6]$ and $CPU_{\text{train}}=1{,}934.1094$ sec for $p_{\text{train}}\in[10/9,7]$.}
\label{tab:fno_p_compare}
\end{table}
Table \ref{tab:fno_p_compare} shows  the $L^\infty$ and $L^2$ inference errors for $p=10/9,\,16/9,\,2,\,3,\,4,\,4.5,\,5,\,7, 9, 11, 25$ after 
training on $p\in[2,6]$ (left) and $p\in[10/9,7]$ with fixed $b=\gamma=1$. 
We observe that the inference error is generally in the order of $O(10^{-1})$, which is worse than the DeepONet. Figure \ref{fig:FNO_epoch} shows that the neural network has already converged, but the solution errors remain large.

Fixing $p_{\text{train}}\in[10/9,7]$, Table \ref{tab:fno_p_compare2}  presents the errors of inferred solutions on $p$ when the FNO is trained  with two $b$ training ranges: a single value $b_{\text{train}}=1$ (left) and  a range $b_{\text{train}}\in[0.5,2.5]$ (right). Table \ref{tab:fno_compare3} lists the errors of inferred solutions on $b$ when the FNO is trained with two $p$ training ranges $p_{\text{train}}\in[10/9,7]$ (left) and a single value $p_{\text{train}}=3$ (right) both with $b_{\text{train}}\in[0.5,2.5]$.  Both tables show the consistent errors and trends as in Table \ref{tab:fno_p_compare}.

{
\begin{table}[h!]
\centering
\setlength{\tabcolsep}{3pt}
\renewcommand{\arraystretch}{1.05}
\begin{tabular}{|c||c|c||c|c|}
\hline
 & \multicolumn{2}{c||}{$b_{\text{train}}=1$} & \multicolumn{2}{c|}{$b_{\text{train}}\in[0.5,2.5]$} \\
\hline\hline
$p$ & $L^\infty$ error & $L^2$ error & $L^\infty$ error & $L^2$ error \\
\hline\hline
$10/9$ & 1.107145e+00 & 8.851508e-01 & 2.897075e+00 & 1.130463e+00 \\
\hline
$16/9$ &  5.876326e-01 & 4.187138e-01 & 1.962677e+00 & 1.050589e+00 \\
\hline
$2$    & 4.741016e-01 & 3.059828e-01 & 1.803749e+00 & 9.820820e-01 \\
\hline
$3$ &  3.232563e-01 & 1.888838e-01 & 8.588895e-01 & 3.049548e-01 \\
\hline
$4$    & 4.777224e-01 & 1.804326e-01 & 5.567601e-01 & 2.031769e-01 \\
\hline
$4.5$  & 5.074022e-01 & 1.919678e-01 & 5.727935e-01 & 1.818893e-01 \\
\hline
$5$    & 5.651524e-01 & 2.097748e-01 & 6.083186e-01 & 1.749442e-01 \\
\hline
$7$    & 7.462914e-01 & 2.453445e-01 & 9.969500e-01 & 4.584269e-01 \\ \hline
9 & 5.504158e-01 & 3.604365e-01 & 1.554534e+00 & 9.600010e-01\\ \hline
11  & 8.165537e-01 & 6.160263e-01 & 2.145100e+00 & 1.535665e+00\\ \hline
25 & 3.186769e+00 & 2.517480e+00 & 6.166720e+00 & 5.735994e+00\\ \hline
\end{tabular}
\caption{FNO performance across $p$ for two $b$ training ranges, with $p_{\text{train}}\in[10/9,7]$, $N=2^{10}$, $L=30$, $W=64$, $L_{FNO}=4$, $B=8$, $lr=5\times 10^{-4}$, and $E=5{,}000$ for $Q''=bQ-Q^p$. Training times: $CPU_{\text{train}}=1{,}934.1094$ sec for $b_{train}=1$ and $1{,}830.1875$ sec for $b_{train}\in[0.5,2.5]$.}
\label{tab:fno_p_compare2}
\end{table}
\begin{table}[h!]
\centering
\setlength{\tabcolsep}{3pt}
\renewcommand{\arraystretch}{1.05}
\begin{tabular}{|c||c|c||c|c|}
\hline
 & \multicolumn{2}{c||}{$p_{\text{train}}\in[10/9,7]$} & \multicolumn{2}{c|}{$p_{\text{train}}=3$} \\
\hline
\hline
$b$ & $L^\infty$ error & $L^2$ error & $L^\infty$ error & $L^2$ error \\
\hline\hline
0.5 & 8.482045e-01 & 2.077942e-01 & 5.703015e-01 & 2.174834e-01 \\
\hline
1.0 & 8.588895e-01 & 3.049548e-01 & 3.485801e-01 & 1.887441e-01 \\
\hline
1.5 & 1.085136e+00 & 5.451924e-01 & 4.072782e-01 & 3.474212e-01 \\
\hline
2.0 & 1.351319e+00 & 7.474299e-01 & 3.441592e-01 & 3.033726e-01 \\
\hline
2.5 & 1.515785e+00 & 7.881066e-01 & 2.078502e-01 & 1.836762e-01 \\
\hline
\end{tabular}
\caption{FNO performance for two $p$ training ranges with $b_{\text{train}}\in[0.5,2.5]$, $N=2^{10}$, $L=30$, $W=64$, $L_{FNO}=4$, $B=8$, $lr=5\times10^{-4}$, and $E=5{,}000$ for $Q''=bQ-Q^3$. Training times: $CPU_{\text{train}}=1830.1875$ sec for $p_{train}\in[10/9,7]$ and $1,754.3906$ sec for $p_{train}=3$.}
\label{tab:fno_compare3}
\end{table}
}

\medskip

Overall, we observed that FNO often underperforms DeepONet in our experiments, which may be due to its grid-dependent Fourier representation with truncated modes. This representation can limit accuracy for localized solutions or ill-conditioned operators. In contrast, DeepONet explicitly factorizes the solution operator and allows mesh-free evaluation, which appears to offer greater flexibility in these settings.

\section{Conclusions and discussion}
In this work, we present a comparison between {\it classical} numerical solvers and {\it neural-network}-based methods for computing ground states or profiles of solitary-wave solutions in a one dimensional setting. Our results confirm that classical approaches retain high-order accuracy and strong computational efficiency for single-instance problems. {\it Physics-informed neural networks} (PINNs) are able to reproduce qualitative solution features but are generally inferior to classical solvers in terms of accuracy and efficiency due to expensive training and slow convergence. {\it Operator-learning} methods exhibit an interesting cost profile: although training is computationally intensive, it can be performed offline and reused across many evaluations. Once trained, these models provide extremely fast, nearly instantaneous inference, making them attractive for applications involving repeated simulations or real-time prediction. For single-instance computations, however, the accuracy of operator-learning methods remains lower than that of classical methods or PINNs in general.  Among the operator-learning approaches considered, DeepONet consistently outperforms FNO.

An interesting direction for future work is the investigation of alternative operator-learning architectures, such as mixture-of-experts (MoE) models \cite{Jacobs1991MOE}, which decompose the solution operator into multiple specialized subnetworks coordinated by a gating mechanism. Such architectures may offer improved robustness and accuracy relative to standard DeepONet formulations, and could further enhance neural solvers for large-scale nonlinear solitary-wave profile computations. In addition, given the localized structure of stationary soliton solutions, future work may explore physics-motivated activation functions (e.g., 
\cite{Sitzmann2020SIREN}) or 
activations inspired by soliton profiles \cite{Boyd2001,Jagtap2020Adaptive}.

\appendix
\renewcommand{\thesection}{\Alph{section}} 

\section{Order reduction induced by Dirichlet boundary conditions}
\label{App:BC}

We provide a formal explanation for the observed $O(h)$ phase error in Table \ref{tab:order_FD2} of \S\ref{sec:FD} when the Dirichlet boundary conditions
\[
u(0)=\sqrt{2}, \qquad u(L)=0
\]
are imposed on the finite interval $[0,L]$.

The original cubic nonlinear infinite--domain problem
\[
u'' = u - u^3, \qquad u'(0)=0, \qquad u(\infty)=0
\]
admits the exact homoclinic solution
\[
u(x)=\sqrt{2}\,\sech x,
\]
which is uniquely determined by the symmetry condition $u'(0)=0$.  
The equation is translation invariant, i.e., there is a  family of homoclinic solutions 
$
u(x)=\sqrt{2}\,\sech(x-\xi),
$
for any $\xi\in\mathbb{R}$, representing a phase translation parameter.

Imposing Dirichlet conditions on a finite interval defines a different boundary value problem, and removes the symmetry constraint $u'(0)=0$ that fixes the phase.  As a result, the three-point finite difference  method with Newton's iteration converges to a discrete solution $u_h$ corresponding to a shifted profile with a small but nonzero phase $\xi_h$. A Taylor expansion near $x=0$ yields
\[
u_h'(0) = -\sqrt{2}\,\sech(\xi_h)\tanh(\xi_h) = O(\xi_h).
\]
In practice, the discrete solution satisfies
$
u_h'(0)=O(h),
$
which implies
$
\xi_h = O(h).
$ This phase error manifests itself as a boundary mismatch.  While the exact solution near $0$, say at $h$, satisfies
\[
u(h)=\sqrt{2}+O(h^2),
\]
the numerical solution behaves as
\[
u_h(h)=\sqrt{2}+u_h'(0)\,h+O(h^2)=\sqrt{2}+O(h).
\]
Thus, an $O(h)$ boundary error is introduced, even when a second--order finite difference scheme is used in the interior. Comparing the numerical solution with the exact profile, we write
\[
u_h(x)=\sqrt{2}\,\sech(x-\xi_h)+O(h^2),
\]
where the $O(h^2)$ term represents the interior discretization error.  
Subtracting $u(x)=\sqrt{2}\,\sech x$ gives
\[
u_h(x)-u(x)=\xi_h\,u'(x)+O(\xi_h^2)+O(h^2).
\]
Since $\xi_h=O(h)$, the phase error dominates, and hence
\[
\|u_h-u\|_\infty = O(h).
\] This is true for $L^2$ norm, as well.

In summary, although the interior discretization is second order and the nonlinear system is solved to high accuracy, the Dirichlet boundary conditions eliminate the symmetry that uniquely selects the homoclinic solution.  The resulting $O(h)$ phase error dominates the global error and leads to a reduction to first--order convergence.

\vfill

\section{Tables for classical methods}
\label{App:2}
{
\begin{table}[H]
\centering
\footnotesize
\begin{subtable}[t]{0.14\textwidth}
\footnotesize
\centering
\begin{tabular}{|c|c|}
\hline
$p$ & $k$ \\ \hline\hline
10/9 & 0.50\\ \hline
10/9 & 0.75\\\hline
10/9 & 0.90\\\hline
10/9 & 1.00\\\hline
10/9 & 1.10\\\hline
10/9 & 1.25\\\hline\hline
16/9 & 0.50\\\hline
16/9 & 0.75\\\hline
16/9 & 0.90\\\hline
16/9 & 1.00\\\hline
16/9 & 1.10\\\hline
16/9 & 1.25\\\hline\hline
2 & 0.50\\\hline
2 & 0.75\\\hline
2 & 0.90\\\hline
2 & 1.00\\\hline
2 & 1.10\\\hline
2 & 1.25\\\hline\hline
3 & 0.50\\\hline
3 & 0.75\\\hline
3 & 0.90\\\hline
3 & 1.00\\\hline
3 & 1.10\\\hline
3 & 1.25\\\hline\hline
4 & 0.50\\\hline
4 & 0.75\\\hline
4 & 0.90\\\hline
4 & 1.00\\\hline
4 & 1.10\\\hline
4 & 1.25\\\hline\hline
4.5 & 0.50\\\hline
4.5 & 0.75\\\hline
4.5 & 0.90\\\hline
4.5 & 1.00\\\hline
4.5 & 1.10\\\hline
4.5 & 1.25\\\hline\hline
5 & 0.50\\\hline
5 & 0.75\\\hline
5 & 0.90\\\hline
5 & 1.00\\\hline
5 & 1.10\\\hline
5 & 1.25\\\hline\hline
7 & 0.50\\\hline
7 & 0.75\\\hline
7 & 0.90\\\hline
7 & 1.00\\\hline
7 & 1.10\\\hline
7 & 1.25\\\hline
\end{tabular}
\end{subtable}
\begin{subtable}[t]{0.42\textwidth}
\footnotesize
\centering
\begin{tabular}{|c|c|c|c|}
\hline
iter & CPU time & $L^\infty$ error & $L^2$ error\\ \hline\hline
 12 & 0.640625 & 1.943589e-03 & 4.408515e-03\\ \hline
 11 & 0.609375 & 1.943589e-03 & 4.408515e-03\\ \hline
 9 & 0.421875 & 1.943589e-03 & 4.408516e-03\\ \hline
0 & - & - & -\\ \hline
 9 & 0.406250 & 1.933107e-03 & 4.385094e-03\\ \hline
 10 & 0.515625 & 1.933108e-03 & 4.385098e-03\\ \hline\hline
 11 & 0.515625 & 4.506001e-03 & 6.400347e-03\\
\hline
\end{tabular}
\caption{FD + Newton}
\end{subtable}
\hfill
\begin{subtable}[t]{0.42\textwidth}
\footnotesize
\centering
\begin{tabular}{|c|c|c|c|}
\hline
 iter & CPU time & $L^\infty$ error & $L^2$ error\\ \hline\hline
 6 & 0.015625 & 2.169142e-08 & 2.278022e-08\\ \hline
 6 & $<10^{-8}$ & 2.169142e-08 & 2.278022e-08\\ \hline
 6 & $<10^{-8}$ & 2.169142e-08 & 2.278022e-08\\ \hline
 0 & - & - & -\\ \hline
 6 & $<10^{-8}$ & 2.169142e-08 & 2.278022e-08\\ \hline
 6 & 0.015625 & 2.169142e-08 & 2.278022e-08\\ \hline\hline
 2 & $<10^{-8}$ & 8.483249e-13 & 8.483566e-13\\
\hline
\end{tabular}
\caption{Petviashvili, factor $\Gamma=\frac{p}{p-1}$}
\end{subtable}
\caption{Comparison of FD+Newton and Petviashvili, \textbf{Sech Variable Amplitude} $U^{(A)}_0=kA \sech^{\frac{2}{p-1}}(\frac{p-1}{2} x)$ with $L=30$, $N_{\text{interior}}=2^{10}$ and tol $=10^{-12}$.} \label{tab:appendix_1}
\end{table}
}
\begin{table}[H]
\centering
\footnotesize
\begin{subtable}[t]{0.14\textwidth}
\footnotesize
\centering
\begin{tabular}{|c|c|}
\hline
$p$ & $k$ \\ \hline\hline
10/9 & 0.50\\ \hline
10/9 & 0.75\\\hline
10/9 & 0.90\\\hline
10/9 & 1.00\\\hline
10/9 & 1.10\\\hline
10/9 & 1.25\\\hline\hline
16/9 & 0.50\\\hline
16/9 & 0.75\\\hline
16/9 & 0.90\\\hline
16/9 & 1.00\\\hline
16/9 & 1.10\\\hline
16/9 & 1.25\\\hline\hline
2 & 0.50\\\hline
2 & 0.75\\\hline
2 & 0.90\\\hline
2 & 1.00\\\hline
2 & 1.10\\\hline
2 & 1.25\\\hline\hline
3 & 0.50\\\hline
3 & 0.75\\\hline
3 & 0.90\\\hline
3 & 1.00\\\hline
3 & 1.10\\\hline
3 & 1.25\\\hline\hline
4 & 0.50\\\hline
4 & 0.75\\\hline
4 & 0.90\\\hline
4 & 1.00\\\hline
4 & 1.10\\\hline
4 & 1.25\\\hline\hline
4.5 & 0.50\\\hline
4.5 & 0.75\\\hline
4.5 & 0.90\\\hline
4.5 & 1.00\\\hline
4.5 & 1.10\\\hline
4.5 & 1.25\\\hline\hline
5 & 0.50\\\hline
5 & 0.75\\\hline
5 & 0.90\\\hline
5 & 1.00\\\hline
5 & 1.10\\\hline
5 & 1.25\\\hline\hline
7 & 0.50\\\hline
7 & 0.75\\\hline
7 & 0.90\\\hline
7 & 1.00\\\hline
7 & 1.10\\\hline
7 & 1.25\\\hline
\end{tabular}
\end{subtable}
\hfill
\begin{subtable}[t]{0.42\textwidth}
\footnotesize
\centering
\begin{tabular}{|c|c|c|c|}
\hline
iter & CPU time & $L^\infty$ error & $L^2$ error\\ \hline\hline
20 & 1.000000 & 1.933106e-03 & 4.385094e-03\\ \hline
14 & 0.656250 & 1.933106e-03 & 4.385094e-03\\ \hline
10 & 0.484375 & 1.933107e-03 & 4.385094e-03\\ \hline
0 & - & - & -\\ \hline
 9 & 0.437500 & 1.943591e-03 & 4.408520e-03\\ \hline
11 & 0.546875 & 1.943589e-03 & 4.408515e-03\\ \hline\hline
15 & 0.734375 & 4.436074e-03 & 6.302829e-03\\
\hline
\end{tabular}
\caption{FD + Newton}
\end{subtable}
\hfill
\begin{subtable}[t]{0.42\textwidth}
\footnotesize
\centering
\begin{tabular}{|c|c|c|c|}
\hline
iter & CPU time & $L^\infty$ error & $L^2$ error\\ \hline\hline
257 & 0.046875 & 2.169161e-08 & 2.278066e-08\\ \hline
244 & 0.046875 & 2.169161e-08 & 2.278066e-08\\ \hline
232 & 0.062500 & 2.169161e-08 & 2.278066e-08\\ \hline
0 & - & - & -\\ \hline
229 & 0.046875 & 2.169161e-08 & 2.278066e-08\\ \hline
236 & 0.046875 & 2.169161e-08 & 2.278066e-08\\ \hline\hline
61 & 0.031250 & 1.483924e-12 & 3.212068e-12\\
\hline
\end{tabular}
\caption{Petviashvili, factor $\Gamma=\frac{p}{p-1}$}
\end{subtable}

\caption{Comparison of FD+Newton and Petviashvili, \textbf{Sech Variable Width} $U^{(W)}_0=A \sech^{\frac{2}{p-1}}(\frac{p-1}{2} k x)$ with $L=30$, $N_{\text{interior}}=2^{10}$ and tol $=10^{-12}$.}
\end{table}


\begin{table}[H]
\centering
\footnotesize

\begin{subtable}[t]{0.14\textwidth}
\footnotesize
\centering
\begin{tabular}{|c|c|}
\hline
$p$ & $k$ \\ \hline\hline
10/9 & 0.50\\ \hline
10/9 & 0.75\\ \hline
10/9 & 0.90\\ \hline
10/9 & 1.00\\ \hline
10/9 & 1.10\\ \hline
10/9 & 1.25\\ \hline\hline
16/9 & 0.50\\ \hline
16/9 & 0.75\\ \hline
16/9 & 0.90\\ \hline
16/9 & 1.00\\ \hline
16/9 & 1.10\\ \hline
16/9 & 1.25\\ \hline\hline
2 & 0.50\\ \hline
2 & 0.75\\ \hline
2 & 0.90\\ \hline
2 & 1.00\\ \hline
2 & 1.10\\ \hline
2 & 1.25\\ \hline\hline
3 & 0.50\\ \hline
3 & 0.75\\ \hline
3 & 0.90\\ \hline
3 & 1.00\\ \hline
3 & 1.10\\ \hline
3 & 1.25\\ \hline\hline
4 & 0.50\\ \hline
4 & 0.75\\ \hline
4 & 0.90\\ \hline
4 & 1.00\\ \hline
4 & 1.10\\ \hline
4 & 1.25\\ \hline\hline
4.5 & 0.50\\ \hline
4.5 & 0.75\\ \hline
4.5 & 0.90\\ \hline
4.5 & 1.00\\ \hline
4.5 & 1.10\\ \hline
4.5 & 1.25\\ \hline\hline
5 & 0.50\\ \hline
5 & 0.75\\ \hline
5 & 0.90\\ \hline
5 & 1.00\\ \hline
5 & 1.10\\ \hline
5 & 1.25\\ \hline\hline
7 & 0.50\\ \hline
7 & 0.75\\ \hline
7 & 0.90\\ \hline
7 & 1.00\\ \hline
7 & 1.10\\ \hline
7 & 1.25\\ \hline
\end{tabular}
\end{subtable}
\hfill
\begin{subtable}[t]{0.42\textwidth}
\footnotesize
\centering
\begin{tabular}{|c|c|c|c|}
\hline
iter & CPU time & $L^\infty$ error & $L^2$ error\\ \hline\hline
16 & 0.781250 & 1.933106e-03 & 4.385094e-03\\ \hline
11 & 0.531250 & 1.933107e-03 & 4.385094e-03\\ \hline
 9 & 0.453125 & 1.933106e-03 & 4.385094e-03\\ \hline
0 & - & - & - \\ \hline
 9 & 0.437500 & 1.943589e-03 & 4.408515e-03\\ \hline
10 & 0.484375 & 1.943589e-03 & 4.408515e-03\\ \hline\hline
13 & 0.656250 & 4.436074e-03 & 6.302829e-03\\
\hline
\end{tabular}
\caption{FD + Newton}
\end{subtable}
\hfill
\begin{subtable}[t]{0.42\textwidth}
\footnotesize
\centering
\begin{tabular}{|c|c|c|c|}
\hline
iter & CPU time & $L^\infty$ error & $L^2$ error\\ \hline\hline
281 & 0.062500 & 2.169161e-08 & 2.278066e-08\\ \hline
277 & 0.046875 & 2.169161e-08 & 2.278066e-08\\ \hline
275 & 0.062500 & 2.169161e-08 & 2.278066e-08\\ \hline
0 & - & - & - \\ \hline
273 & 0.062500 & 2.169161e-08 & 2.278066e-08\\ \hline
272 & 0.062500 & 2.169161e-08 & 2.278066e-08\\ \hline\hline
62 & 0.015625 & 1.594280e-12 & 3.435759e-12\\
\hline
\end{tabular}
\caption{Petviashvili, factor $\Gamma=\frac{p}{p-1}$}
\end{subtable}
\caption{Comparison of FD+Newton and Petviashvili, \textbf{Sech Variable Power} $U^{(P)}_0=A \sech^{\frac{2k}{p-1}}(\frac{p-1}{2}x)$ with $L=30$, $N_{\text{interior}}=2^{10}$ and tol $=10^{-12}$.}
\end{table}

\begin{table}[H]
\centering
\footnotesize
\begin{subtable}[t]{0.14\textwidth}
\footnotesize
\centering
\begin{tabular}{|c|c|}
\hline
$p$ & $k$ \\ \hline\hline
10/9 & 0.50\\ \hline
10/9 & 0.75\\ \hline
10/9 & 0.90\\ \hline
10/9 & 1.00\\ \hline
10/9 & 1.10\\ \hline
10/9 & 1.25\\ \hline\hline
16/9 & 0.50\\ \hline
16/9 & 0.75\\ \hline
16/9 & 0.90\\ \hline
16/9 & 1.00\\ \hline
16/9 & 1.10\\ \hline
16/9 & 1.25\\ \hline\hline
2 & 0.50\\ \hline
2 & 0.75\\ \hline
2 & 0.90\\ \hline
2 & 1.00\\ \hline
2 & 1.10\\ \hline
2 & 1.25\\ \hline\hline
3 & 0.50\\ \hline
3 & 0.75\\ \hline
3 & 0.90\\ \hline
3 & 1.00\\ \hline
3 & 1.10\\ \hline
3 & 1.25\\ \hline\hline
4 & 0.50\\ \hline
4 & 0.75\\ \hline
4 & 0.90\\ \hline
4 & 1.00\\ \hline
4 & 1.10\\ \hline
4 & 1.25\\ \hline\hline
4.5 & 0.50\\ \hline
4.5 & 0.75\\ \hline
4.5 & 0.90\\ \hline
4.5 & 1.00\\ \hline
4.5 & 1.10\\ \hline
4.5 & 1.25\\ \hline\hline
5 & 0.50\\ \hline
5 & 0.75\\ \hline
5 & 0.90\\ \hline
5 & 1.00\\ \hline
5 & 1.10\\ \hline
5 & 1.25\\ \hline\hline
7 & 0.50\\ \hline
7 & 0.75\\ \hline
7 & 0.90\\ \hline
7 & 1.00\\ \hline
7 & 1.10\\ \hline
7 & 1.25\\ \hline
\end{tabular}
\end{subtable}
\hfill
\begin{subtable}[t]{0.42\textwidth}
\footnotesize
\centering
\begin{tabular}{|c|c|c|c|}
\hline
iter & CPU time & $L^\infty$ error & $L^2$ error\\ \hline\hline
14 & 0.656250 & 1.943589e-03 & 4.408515e-03\\ \hline
14 & 0.671875 & 1.943589e-03 & 4.408515e-03\\ \hline
14 & 0.687500 & 1.943589e-03 & 4.408515e-03\\ \hline
14 & 0.671875 & 1.943589e-03 & 4.408515e-03\\ \hline
14 & 0.671875 & 1.943589e-03 & 4.408515e-03\\ \hline
14 & 0.671875 & 1.943589e-03 & 4.408515e-03\\ \hline\hline
11 & 0.5,00000 & 4.506001e-03 & 6.400347e-03\\
\hline
\end{tabular}
\caption{FD + Newton}
\end{subtable}
\hfill
\begin{subtable}[t]{0.42\textwidth}
\footnotesize
\centering
\begin{tabular}{|c|c|c|c|}
\hline
iter & CPU time & $L^\infty$ error & $L^2$ error\\ \hline\hline
247 & 0.046875 & 2.169161e-08 & 2.278066e-08\\ \hline
247 & 0.046875 & 2.169161e-08 & 2.278066e-08\\ \hline
247 & 0.046875 & 2.169161e-08 & 2.278066e-08\\ \hline
247 & 0.046875 & 2.169161e-08 & 2.278066e-08\\ \hline
247 & 0.046875 & 2.169161e-08 & 2.278066e-08\\ \hline
247 & 0.046875 & 2.169161e-08 & 2.278066e-08\\ \hline\hline
59 & 0.031250 & 1.377565e-12 & 2.999007e-12\\
\hline
\end{tabular}
\caption{Petviashvili, factor $\Gamma=\frac{p}{p-1}$}
\end{subtable}
\caption{Comparison of FD+Newton and Petviashvili, \textbf{Gaussian Variable Amplitude} $U^{(A)}_{0}=kAe^{-x^2/A}$ with $L=30$, $N_{\text{interior}}=2^{10}$ and tol $=10^{-12}$.}
\end{table}

\begin{table}[H]
\centering
\footnotesize
\begin{subtable}[t]{0.14\textwidth}
\footnotesize
\centering
\begin{tabular}{|c|c|}
\hline
$p$ & $k$ \\ \hline\hline
10/9 & 0.50\\ \hline
10/9 & 0.75\\ \hline
10/9 & 0.90\\ \hline
10/9 & 1.00\\ \hline
10/9 & 1.10\\ \hline
10/9 & 1.25\\ \hline\hline
16/9 & 0.50\\ \hline
16/9 & 0.75\\ \hline
16/9 & 0.90\\ \hline
16/9 & 1.00\\ \hline
16/9 & 1.10\\ \hline
16/9 & 1.25\\ \hline\hline
2 & 0.50\\ \hline
2 & 0.75\\ \hline
2 & 0.90\\ \hline
2 & 1.00\\ \hline
2 & 1.10\\ \hline
2 & 1.25\\ \hline\hline
3 & 0.50\\ \hline
3 & 0.75\\ \hline
3 & 0.90\\ \hline
3 & 1.00\\ \hline
3 & 1.10\\ \hline
3 & 1.25\\ \hline\hline
4 & 0.50\\ \hline
4 & 0.75\\ \hline
4 & 0.90\\ \hline
4 & 1.00\\ \hline
4 & 1.10\\ \hline
4 & 1.25\\ \hline\hline
4.5 & 0.50\\ \hline
4.5 & 0.75\\ \hline
4.5 & 0.90\\ \hline
4.5 & 1.00\\ \hline
4.5 & 1.10\\ \hline
4.5 & 1.25\\ \hline\hline
5 & 0.50\\ \hline
5 & 0.75\\ \hline
5 & 0.90\\ \hline
5 & 1.00\\ \hline
5 & 1.10\\ \hline
5 & 1.25\\ \hline\hline
7 & 0.50\\ \hline
7 & 0.75\\ \hline
7 & 0.90\\ \hline
7 & 1.00\\ \hline
7 & 1.10\\ \hline
7 & 1.25\\ \hline
\end{tabular}
\end{subtable}
\hfill
\begin{subtable}[t]{0.42\textwidth}
\footnotesize
\centering
\begin{tabular}{|c|c|c|c|}
\hline
iter & CPU time & $L^\infty$ error & $L^2$ error\\ \hline\hline
13 & 0.609375 & 1.943589e-03 & 4.408516e-03\\ \hline
14 & 0.671875 & 1.943589e-03 & 4.408515e-03\\ \hline
14 & 0.671875 & 1.943589e-03 & 4.408515e-03\\ \hline
14 & 0.671875 & 1.943589e-03 & 4.408515e-03\\ \hline
14 & 0.656250 & 1.943589e-03 & 4.408515e-03\\ \hline
14 & 0.671875 & 1.943589e-03 & 4.408515e-03\\ \hline\hline
10 & 0.484375 & 4.506001e-03 & 6.400347e-03\\
\hline
\end{tabular}
\caption{FD + Newton}
\end{subtable}
\hfill
\begin{subtable}[t]{0.42\textwidth}
\footnotesize
\centering
\begin{tabular}{|c|c|c|c|}
\hline
iter & CPU time & $L^\infty$ error & $L^2$ error\\ \hline\hline
246 & 0.046875 & 2.169161e-08 & 2.278066e-08\\ \hline
246 & 0.046875 & 2.169161e-08 & 2.278066e-08\\ \hline
247 & 0.046875 & 2.169161e-08 & 2.278066e-08\\ \hline
247 & 0.046875 & 2.169161e-08 & 2.278066e-08\\ \hline
247 & 0.046875 & 2.169161e-08 & 2.278066e-08\\ \hline
247 & 0.046875 & 2.169161e-08 & 2.278066e-08\\ \hline\hline
58 & 0.031250 & 1.295852e-12 & 2.835922e-12\\
\hline
\end{tabular}
\caption{Petviashvili, factor $\Gamma=\frac{p}{p-1}$}
\end{subtable}
\caption{Comparison of FD+Newton and Petviashvili, \textbf{Gaussian Variable Width} $U^{(W)}_{0}=A e^{-kx^2/A}$ with $L=30$, $N_{\text{interior}}=2^{10}$ and tol $=10^{-12}$.}
\end{table}

\begin{table}[H]
\centering
\footnotesize
\begin{subtable}[t]{0.13\textwidth}
\footnotesize
\centering
\begin{tabular}{|c|c|}
\hline
$p$ & $k$ \\ \hline\hline
10/9 & 0.50\\ \hline
10/9 & 0.75\\ \hline
10/9 & 0.90\\ \hline
10/9 & 1.00\\ \hline
10/9 & 1.10\\ \hline
10/9 & 1.25\\ \hline\hline
16/9 & 0.50\\ \hline
16/9 & 0.75\\ \hline
16/9 & 0.90\\ \hline
16/9 & 1.00\\ \hline
16/9 & 1.10\\ \hline
16/9 & 1.25\\ \hline\hline
2 & 0.50\\ \hline
2 & 0.75\\ \hline
2 & 0.90\\ \hline
2 & 1.00\\ \hline
2 & 1.10\\ \hline
2 & 1.25\\ \hline\hline
3 & 0.50\\ \hline
3 & 0.75\\ \hline
3 & 0.90\\ \hline
3 & 1.00\\ \hline
3 & 1.10\\ \hline
3 & 1.25\\ \hline\hline
4 & 0.50\\ \hline
4 & 0.75\\ \hline
4 & 0.90\\ \hline
4 & 1.00\\ \hline
4 & 1.10\\ \hline
4 & 1.25\\ \hline\hline
4.5 & 0.50\\ \hline
4.5 & 0.75\\ \hline
4.5 & 0.90\\ \hline
4.5 & 1.00\\ \hline
4.5 & 1.10\\ \hline
4.5 & 1.25\\ \hline\hline
5 & 0.50\\ \hline
5 & 0.75\\ \hline
5 & 0.90\\ \hline
5 & 1.00\\ \hline
5 & 1.10\\ \hline
5 & 1.25\\ \hline\hline
7 & 0.50\\ \hline
7 & 0.75\\ \hline
7 & 0.90\\ \hline
7 & 1.00\\ \hline
7 & 1.10\\ \hline
7 & 1.25\\ \hline
\end{tabular}
\end{subtable}
\hfill
\begin{subtable}[t]{0.42\textwidth}
\centering
\footnotesize
\begin{tabular}{|c|c|c|c|}
\hline
iter & CPU time & $L^\infty$ error & $L^2$ error\\ \hline\hline
14 & 0.671875 & 1.943589e-03 & 4.408515e-03\\ \hline
14 & 0.703125 & 1.943589e-03 & 4.408515e-03\\ \hline
14 & 0.671875 & 1.943589e-03 & 4.408515e-03\\ \hline
14 & 0.671875 & 1.943589e-03 & 4.408515e-03\\ \hline
14 & 0.671875 & 1.943589e-03 & 4.408515e-03\\ \hline
14 & 0.640625 & 1.943589e-03 & 4.408515e-03\\ \hline\hline
12 & 0.562500 & 4.506001e-03 & 6.400347e-03\\
\hline
\end{tabular}
\caption{FD + Newton}
\end{subtable}
\hfill
\begin{subtable}[t]{0.42\textwidth}
\centering
\footnotesize
\begin{tabular}{|c|c|c|c|}
\hline
iter & CPU time & $L^\infty$ error & $L^2$ error\\ \hline\hline
247 & 0.109375 & 2.169161e-08 & 2.277864e-08\\ \hline
247 & 0.109375 & 2.169161e-08 & 2.277864e-08\\ \hline
247 & 0.093750 & 2.169161e-08 & 2.277864e-08\\ \hline
247 & 0.109375 & 2.169161e-08 & 2.277864e-08\\ \hline
247 & 0.109375 & 2.169161e-08 & 2.277864e-08\\ \hline
247 & 0.109375 & 2.169161e-08 & 2.277864e-08\\ \hline\hline
59 & 0.031250 & 1.498579e-12 & 3.243349e-12\\
\hline
\end{tabular}
\caption{Petviashvili, factor $\Gamma=\frac{p}{p-1}$}
\end{subtable}
\caption{Comparison of FD+Newton and Petviashvili, \textbf{Super Gaussian Variable Amplitude} 
$U^{(A)}_0 = kA e^{-x^4/A}$ with $L=30$, $N_{\text{interior}}=2^{10}$, tol $=10^{-12}$.}
\end{table}

\begin{table}[H]
\centering
\footnotesize
\begin{subtable}[t]{0.14\textwidth}
\footnotesize
\centering
\begin{tabular}{|c|c|}
\hline
$p$ & $k$ \\ \hline\hline
10/9 & 0.50\\ \hline
10/9 & 0.75\\ \hline
10/9 & 0.90\\ \hline
10/9 & 1.00\\ \hline
10/9 & 1.10\\ \hline
10/9 & 1.25\\ \hline\hline
16/9 & 0.50\\ \hline
16/9 & 0.75\\ \hline
16/9 & 0.90\\ \hline
16/9 & 1.00\\ \hline
16/9 & 1.10\\ \hline
16/9 & 1.25\\ \hline\hline
2 & 0.50\\ \hline
2 & 0.75\\ \hline
2 & 0.90\\ \hline
2 & 1.00\\ \hline
2 & 1.10\\ \hline
2 & 1.25\\ \hline\hline
3 & 0.50\\ \hline
3 & 0.75\\ \hline
3 & 0.90\\ \hline
3 & 1.00\\ \hline
3 & 1.10\\ \hline
3 & 1.25\\ \hline\hline
4 & 0.50\\ \hline
4 & 0.75\\ \hline
4 & 0.90\\ \hline
4 & 1.00\\ \hline
4 & 1.10\\ \hline
4 & 1.25\\ \hline\hline
4.5 & 0.50\\ \hline
4.5 & 0.75\\ \hline
4.5 & 0.90\\ \hline
4.5 & 1.00\\ \hline
4.5 & 1.10\\ \hline
4.5 & 1.25\\ \hline\hline
5 & 0.50\\ \hline
5 & 0.75\\ \hline
5 & 0.90\\ \hline
5 & 1.00\\ \hline
5 & 1.10\\ \hline
5 & 1.25\\ \hline\hline
7 & 0.50\\ \hline
7 & 0.75\\ \hline
7 & 0.90\\ \hline
7 & 1.00\\ \hline
7 & 1.10\\ \hline
7 & 1.25\\ \hline
\end{tabular}
\end{subtable}
\hfill
\begin{subtable}[t]{0.42\textwidth}
\centering
\footnotesize
\begin{tabular}{|c|c|c|c|}
\hline
iter & CPU time & $L^\infty$ error & $L^2$ error\\ \hline\hline
14 & 0.703125 & 1.943589e-03 & 4.408515e-03\\ \hline
14 & 0.671875 & 1.943589e-03 & 4.408515e-03\\ \hline
14 & 0.671875 & 1.943589e-03 & 4.408515e-03\\ \hline
14 & 0.687500 & 1.943589e-03 & 4.408515e-03\\ \hline
14 & 0.687500 & 1.943589e-03 & 4.408515e-03\\ \hline
14 & 0.671875 & 1.943589e-03 & 4.408515e-03\\ \hline\hline
11 & 0.531250 & 4.506001e-03 & 6.400347e-03\\
\hline
\end{tabular}
\caption{FD + Newton}
\end{subtable}
\hfill
\begin{subtable}[t]{0.42\textwidth}
\centering
\footnotesize
\begin{tabular}{|c|c|c|c|}
\hline
iter & CPU time & $L^\infty$ error & $L^2$ error\\ \hline\hline
247 & 0.109375 & 2.169161e-08 & 2.277864e-08\\ \hline
247 & 0.109375 & 2.169161e-08 & 2.277864e-08\\ \hline
247 & 0.093750 & 2.169161e-08 & 2.277864e-08\\ \hline
247 & 0.109375 & 2.169161e-08 & 2.277864e-08\\ \hline
247 & 0.093750 & 2.169161e-08 & 2.277864e-08\\ \hline
247 & 0.093750 & 2.169161e-08 & 2.277864e-08\\ \hline\hline
59 & 0.031250 & 1.279088e-12 & 2.802851e-12\\
\hline
\end{tabular}
\caption{Petviashvili, factor $\Gamma=\frac{p}{p-1}$}
\end{subtable}
\caption{Comparison of FD+Newton and Petviashvili, \textbf{Super Gaussian Variable Width} $U^{(W)}_0=A e^{-kx^4/A}$ with $L=30$, $N_{\text{interior}}=2^{10}$ and tol $=10^{-12}$.}
\end{table}


\begin{table}[H]
\centering
\footnotesize
\begin{subtable}[t]{0.12\textwidth}
\centering
\begin{tabular}{|c|c|}
\hline
$p$ & $k$ \\ \hline\hline
10/9 & 0.50\\ \hline
10/9 & 0.75\\ \hline
10/9 & 0.90\\ \hline
10/9 & 1.00\\ \hline
10/9 & 1.10\\ \hline
10/9 & 1.25\\ \hline\hline
16/9 & 0.50\\ \hline
16/9 & 0.75\\ \hline
16/9 & 0.90\\ \hline
16/9 & 1.00\\ \hline
16/9 & 1.10\\ \hline
16/9 & 1.25\\ \hline\hline
2 & 0.50\\ \hline
2 & 0.75\\ \hline
2 & 0.90\\ \hline
2 & 1.00\\ \hline
2 & 1.10\\ \hline
2 & 1.25\\ \hline\hline
3 & 0.50\\ \hline
3 & 0.75\\ \hline
3 & 0.90\\ \hline
3 & 1.00\\ \hline
3 & 1.10\\ \hline
3 & 1.25\\ \hline\hline
4 & 0.50\\ \hline
4 & 0.75\\ \hline
4 & 0.90\\ \hline
4 & 1.00\\ \hline
4 & 1.10\\ \hline
4 & 1.25\\ \hline\hline
4.5 & 0.50\\ \hline
4.5 & 0.75\\ \hline
4.5 & 0.90\\ \hline
4.5 & 1.00\\ \hline
4.5 & 1.10\\ \hline
4.5 & 1.25\\ \hline\hline
5 & 0.50\\ \hline
5 & 0.75\\ \hline
5 & 0.90\\ \hline
5 & 1.00\\ \hline
5 & 1.10\\ \hline
5 & 1.25\\ \hline\hline
7 & 0.50\\ \hline
7 & 0.75\\ \hline
7 & 0.90\\ \hline
7 & 1.00\\ \hline
7 & 1.10\\ \hline
7 & 1.25\\ \hline
\end{tabular}
\end{subtable}
\hfill
\begin{subtable}[t]{0.43\textwidth}
\centering
\begin{tabular}{|c|c|c|c|}
\hline
iter & CPU time & $L^\infty$ error & $L^2$ error\\ \hline\hline
 265 & 0.062500 & 2.169161e-08 & 2.278066e-08\\ \hline
 265 & 0.046875 & 2.169161e-08 & 2.278066e-08\\ \hline
 265 & 0.062500 & 2.169161e-08 & 2.278066e-08\\ \hline
 265 & 0.062500 & 2.169161e-08 & 2.278066e-08\\ \hline
 265 & 0.046875 & 2.169161e-08 & 2.278066e-08\\ \hline
 265 & 0.062500 & 2.169161e-08 & 2.278066e-08\\ \hline\hline
 65 & 0.015625 & 1.636358e-12 & 3.520168e-12\\
\hline
\end{tabular}
\caption{Hat Function Variable Amplitude}
\end{subtable}
\hfill
\begin{subtable}[t]{0.43\textwidth}
\centering
\begin{tabular}{|c|c|c|c|}
\hline
iter & CPU time & $L^\infty$ error & $L^2$ error\\ \hline\hline
 276 & 0.062500 & 2.169161e-08 & 2.278066e-08\\ \hline
 270 & 0.046875 & 2.169161e-08 & 2.278066e-08\\ \hline
 267 & 0.062500 & 2.169161e-08 & 2.278066e-08\\ \hline
 265 & 0.046875 & 2.169161e-08 & 2.278066e-08\\ \hline
 264 & 0.046875 & 2.169161e-08 & 2.278066e-08\\ \hline
 261 & 0.062500 & 2.169161e-08 & 2.278066e-08\\ \hline\hline
 67 & 0.015625 & 1.389555e-12 & 3.021678e-12\\
\hline
\end{tabular}
\caption{Hat Function Variable Width}
\end{subtable}
\caption{Comparison of Petviashvili iterations for (A) Hat Function Variable Amplitude $U^{(A)}_0 = kA\,\chi_{|x|\leq L}\left(1-\frac{x}{L}\right)$ and (B) Hat Function Variable Width $U^{(B)}_0 = A\,\chi_{|x|\leq \frac{L}{k}}\left(1-\frac{kx}{L}\right)$, with $\Gamma=\frac{p}{p-1}$, $L=30$, $N_{\text{interior}}=2^{10}$ and tol $=10^{-12}$.} 
\label{tab:appendix_end}
\end{table}

\section{Tables for PINN}
\label{App:3}
{
\begin{table}[H]
\centering
\scriptsize
\setlength{\tabcolsep}{2pt}
\renewcommand{\arraystretch}{1.05}
\begin{subtable}[t]{0.14\textwidth}
\centering
\begin{tabular}{|c|c|}
\hline
$p$ & Epochs \\ \hline\hline
10/9 & 5,000\\ \hline
10/9 & 10,000\\ \hline
10/9 & 15,000\\ \hline
10/9 & 20,000\\ \hline
10/9 & 25,000\\ \hline
10/9 & 30,000\\ \hline\hline
16/9 & 5,000\\ \hline
16/9 & 10,000\\ \hline
16/9 & 15,000\\ \hline
16/9 & 20,000\\ \hline
16/9 & 25,000\\ \hline
16/9 & 30,000\\ \hline\hline
2 & 5,000\\ \hline
2 & 10,000\\ \hline
2 & 15,000\\ \hline
2 & 20,000\\ \hline
2 & 25,000\\ \hline
2 & 30,000\\ \hline\hline
3 & 5,000\\ \hline
3 & 10,000\\ \hline
3 & 15,000\\ \hline
3 & 20,000\\ \hline
3 & 25,000\\ \hline
3 & 30,000\\ \hline\hline
4 & 5,000\\ \hline
4 & 10,000\\ \hline
4 & 15,000\\ \hline
4 & 20,000\\ \hline
4 & 25,000\\ \hline
4 & 30,000\\ \hline\hline
4.5 & 5,000\\ \hline
4.5 & 10,000\\ \hline
4.5 & 15,000\\ \hline
4.5 & 20,000\\ \hline
4.5 & 25,000\\ \hline
4.5 & 30,000\\ \hline\hline
5 & 5,000\\ \hline
5 & 10,000\\ \hline
5 & 15,000\\ \hline
5 & 20,000\\ \hline
5 & 25,000\\ \hline
5 & 30,000\\ \hline\hline
7 & 5,000\\ \hline
7 & 10,000\\ \hline
7 & 15,000\\ \hline
7 & 20,000\\ \hline
7 & 25,000\\ \hline
7 & 30,000\\ \hline
\end{tabular}
\end{subtable}
\hfill
\begin{subtable}[t]{0.42\textwidth}
\centering
\begin{tabular}{|c|c|c|c|}
\hline
Loss & CPU time & $L^\infty$ error & $L^2$ error\\ \hline\hline
5.497023e-06 & 319.1562 & 2.243199e-01 & 1.573422e-01 \\  \hline
6.399824e-06 & 640.3906 & 1.453902e-01 & 1.040050e-01 \\  \hline
2.379962e-05 & 942.4688 & 8.503551e-02 & 6.114503e-02 \\  \hline
2.122101e-06 & 1228.2500 & 5.233342e-02 & 3.753107e-02 \\  \hline
7.842139e-07 & 1521.5,000 & 3.784282e-02 & 2.709129e-02 \\  \hline
2.801825e-07 & 1817.3906 & 2.905399e-02 & 2.073294e-02 \\  \hline\hline
4.539448e-06 & 265.7656 & 6.724527e-02 & 5.124874e-02 \\  \hline
6.483661e-05 & 556.9219 & 4.858007e-02 & 3.856062e-02 \\  \hline
3.321990e-06 & 869.5938 & 3.500601e-02 & 2.664079e-02 \\  \hline
2.217234e-06 & 1160.2969 & 2.413239e-02 & 1.869404e-02 \\  \hline
1.906982e-07 & 1440.1250 & 1.583989e-02 & 1.206024e-02 \\  \hline
1.162162e-07 & 1718.0938 & 1.407494e-02 & 1.070179e-02 \\  \hline\hline
5.728005e-06 & 271.7500 & 5.880260e-02 & 4.567062e-02 \\  \hline
1.860798e-06 & 548.1562 & 3.433431e-02 & 2.667928e-02 \\  \hline
6.500152e-05 & 823.7969 & 3.181619e-02 & 2.801086e-02 \\  \hline
2.678400e-05 & 1098.2812 & 2.115374e-02 & 1.792525e-02 \\  \hline
1.377340e-07 & 1371.5938 & 1.360226e-02 & 1.053814e-02 \\  \hline
1.401239e-05 & 1644.0781 & 9.877680e-03 & 8.691553e-03 \\  \hline\hline
3.342220e-06 & 276.1406 & 2.897986e-02 & 2.384620e-02 \\  \hline
7.665640e-07 & 554.4844 & 1.717105e-02 & 1.404423e-02 \\  \hline
1.931678e-07 & 836.4219 & 1.319679e-02 & 1.081650e-02 \\  \hline
9.857257e-07 & 1081.4531 & 8.312488e-03 & 6.801640e-03 \\  \hline
8.122560e-06 & 1326.8750 & 9.003416e-03 & 8.869082e-03 \\  \hline
1.216845e-05 & 1576.0000 & 8.875633e-03 & 7.784941e-03 \\  \hline\hline
5.830221e-06 & 250.5625 & 1.086501e-02 & 9.729904e-03 \\  \hline
1.999822e-05 & 499.1094 & 2.211286e-03 & 6.838305e-03 \\  \hline
2.967012e-07 & 750.7812 & 1.210080e-02 & 1.030578e-02 \\  \hline
9.390449e-06 & 1001.2969 & 1.022211e-02 & 9.032625e-03 \\  \hline
3.043750e-07 & 1256.2031 & 7.990463e-03 & 6.832259e-03 \\  \hline
4.361955e-08 & 1508.5625 & 6.420298e-03 & 5.453788e-03 \\  \hline\hline
5.155213e-06 & 278.2188 & 1.949269e-03 & 3.679638e-03 \\  \hline
5.087513e-07 & 554.1562 & 1.222192e-03 & 1.747193e-03 \\  \hline
9.071521e-06 & 809.5,000 & 1.174661e-02 & 1.024465e-02 \\  \hline
5.820977e-08 & 1080.7188 & 1.012578e-02 & 8.719509e-03 \\  \hline
4.662532e-07 & 1385.8750 & 7.715193e-03 & 6.701888e-03 \\  \hline
1.761227e-07 & 1690.2812 & 6.465403e-03 & 5.551418e-03 \\  \hline\hline
4.656524e-06 & 269.7656 & 2.221244e-03 & 4.426311e-03 \\  \hline
3.944043e-07 & 529.5156 & 7.300320e-03 & 6.346104e-03 \\  \hline
1.284419e-07 & 791.5469 & 1.219529e-02 & 1.067560e-02 \\  \hline
2.939521e-07 & 1052.1875 & 9.939911e-03 & 8.830281e-03 \\  \hline
6.253616e-07 & 1311.0938 & 6.527724e-03 & 5.659281e-03 \\  \hline
8.773534e-08 & 1572.6562 & 5.856099e-03 & 5.055659e-03 \\  \hline\hline
2.168570e-06 & 266.3750 & 2.615644e-03 & 4.677767e-03 \\  \hline
2.552035e-06 & 530.7812 & 1.237343e-02 & 1.177902e-02 \\  \hline
3.781839e-07 & 795.3750 & 1.298075e-02 & 1.196035e-02 \\  \hline
1.979604e-07 & 1064.3125 & 8.607195e-03 & 7.784629e-03 \\  \hline
5.636335e-08 & 1337.6875 & 7.221923e-03 & 6.567768e-03 \\  \hline
6.893825e-08 & 1642.9531 & 6.049440e-03 & 5.461686e-03 \\  \hline
\end{tabular}
\caption{PINN with $\tanh$ activation}
\end{subtable}
\hfill
\begin{subtable}[t]{0.42\textwidth}
\centering
\begin{tabular}{|c|c|c|c|}
\hline
Loss & CPU time & $L^\infty$ error & $L^2$ error\\ \hline\hline
3.125369e-04 & 440.6562 & 8.966974e-02 & 6.164387e-02 \\  \hline
5.098428e-05 & 880.1875 & 5.684752e-02 & 3.972772e-02 \\  \hline
1.946795e-05 & 1341.2656 & 3.135775e-02 & 2.225004e-02 \\  \hline
4.129257e-07 & 1723.9375 & 2.007288e-02 & 1.437145e-02 \\  \hline
2.286919e-07 & 2107.3438 & 1.595683e-02 & 1.144018e-02 \\  \hline
5.874724e-08 & 2495.7969 & 1.405067e-02 & 1.008419e-02 \\  \hline\hline
4.401185e-06 & 395.7969 & 9.267625e-02 & 6.986855e-02 \\  \hline
9.104569e-07 & 784.9219 & 4.159047e-02 & 3.164112e-02 \\  \hline
1.987755e-06 & 1163.7969 & 3.302970e-02 & 2.517328e-02 \\  \hline
1.458786e-07 & 1543.4219 & 2.671899e-02 & 2.037609e-02 \\  \hline
3.227341e-08 & 1930.1406 & 2.376266e-02 & 1.810152e-02 \\  \hline
1.612241e-07 & 2331.1094 & 2.169832e-02 & 1.651699e-02 \\  \hline\hline
2.368115e-06 & 369.4844 & 7.764238e-02 & 5.988062e-02 \\  \hline
4.601865e-06 & 744.0625 & 3.778798e-02 & 2.921325e-02 \\  \hline
1.466359e-07 & 1118.0469 & 3.014737e-02 & 2.335713e-02 \\  \hline
4.041756e-07 & 1522.2031 & 2.524209e-02 & 1.960254e-02 \\  \hline
1.687374e-06 & 1898.1250 & 2.251036e-02 & 1.736689e-02 \\  \hline
3.810378e-08 & 2273.3438 & 2.099043e-02 & 1.627930e-02 \\  \hline\hline
7.009239e-06 & 375.8594 & 5.210724e-02 & 4.298015e-02 \\  \hline
3.027944e-06 & 764.8906 & 3.608915e-02 & 2.985622e-02 \\  \hline
8.476383e-07 & 1175.2031 & 3.082486e-02 & 2.533770e-02 \\  \hline
6.584735e-07 & 1580.4688 & 2.642485e-02 & 2.179384e-02 \\  \hline
5.086789e-07 & 1984.3281 & 2.410283e-02 & 1.985120e-02 \\  \hline
3.858329e-07 & 2388.8594 & 2.076610e-02 & 1.708686e-02 \\  \hline\hline
2.950854e-06 & 405.1094 & 3.499691e-02 & 3.020483e-02 \\  \hline
1.483778e-06 & 782.0312 & 2.467643e-02 & 2.117759e-02 \\  \hline
1.816548e-04 & 1159.0625 & 2.232925e-02 & 2.422558e-02 \\  \hline
4.160280e-07 & 1535.7500 & 1.765202e-02 & 1.504776e-02 \\  \hline
3.545017e-07 & 1937.1719 & 1.767938e-02 & 1.511414e-02 \\  \hline
3.072228e-07 & 2318.4688 & 1.535378e-02 & 1.312327e-02 \\  \hline\hline
2.278833e-06 & 384.2188 & 3.105947e-02 & 2.727231e-02 \\  \hline
6.813670e-07 & 765.1719 & 2.008453e-02 & 1.749996e-02 \\  \hline
6.375276e-07 & 1167.3594 & 1.703819e-02 & 1.500170e-02 \\  \hline
3.517839e-06 & 1574.8906 & 1.593140e-02 & 1.387310e-02 \\  \hline
1.596826e-06 & 1956.9844 & 1.612541e-02 & 1.404605e-02 \\  \hline
4.239125e-07 & 2343.9375 & 1.436826e-02 & 1.242997e-02 \\  \hline\hline
1.936395e-06 & 387.8281 & 2.884829e-02 & 2.567385e-02 \\  \hline
1.896649e-05 & 777.4375 & 1.746059e-02 & 1.559951e-02 \\  \hline
2.202428e-04 & 1178.1250 & 2.059728e-02 & 2.068189e-02 \\  \hline
4.563825e-07 & 1582.9375 & 1.439053e-02 & 1.258496e-02 \\  \hline
1.240757e-06 & 1973.8750 & 1.464433e-02 & 1.286044e-02 \\  \hline
3.601354e-07 & 2363.6719 & 1.315570e-02 & 1.158138e-02 \\  \hline\hline
5.384235e-06 & 392.7969 & 2.118145e-02 & 1.959180e-02 \\  \hline
1.064684e-06 & 800.1875 & 1.618926e-02 & 1.496021e-02 \\  \hline
3.315556e-05 & 1186.6875 & 1.173239e-02 & 1.358119e-02 \\  \hline
3.923431e-07 & 1575.4531 & 9.591446e-03 & 8.759088e-03 \\  \hline
2.989762e-06 & 1959.6094 & 7.116364e-03 & 6.514945e-03 \\  \hline
8.585262e-07 & 2376.9844 & 7.738219e-03 & 7.052342e-03 \\  \hline
\end{tabular}
\caption{PINN with SiLU activation}
\end{subtable}
\caption{Comparison of PINN performance with $\tanh$ and SiLU activations, $N_{f}=2^{10}$, $L=30$, 4 hidden layers and 64 neurons for $Q'' = Q - Q^p$.}
\label{tab:pinn_p}
\end{table}
}

\newpage

\bibliographystyle{acm} 
\bibliography{ref}

\end{document}